\documentclass[aps,prx,twocolumn,showpacs,superscriptaddress,floatfix
,nofootinbib]{revtex4-2}
\usepackage[T1]{fontenc}
\usepackage[utf8]{inputenc}
\usepackage{mathtools}
\usepackage{graphicx}
\usepackage{xcolor}
\usepackage{soul}
\usepackage{amsmath,amssymb}
\usepackage[colorlinks,citecolor=blue,linkcolor=blue,anchorcolor=blue,filecolor=blue, urlcolor=blue]{hyperref}
\usepackage{orcidlink}
\usepackage{lipsum}

\usepackage{ae,aecompl}
\usepackage{hyperref}
\usepackage{amsmath}
\usepackage{amssymb}
\usepackage{bm}
\usepackage{cleveref}
\usepackage{url}

\begin{document}

\author{Adamu Issifu \orcidlink{0000-0002-2843-835X}} 
\email{ai@academico.ufpb.br}

\affiliation{Instituto Tecnológico de Aeronáutica,\\ CEP 12.228-900, São José dos Campos, SP, Brazil} 

\author{D\'ebora P. Menezes \orcidlink{0000-0003-0730-6689}}
\email{debora.p.m@ufsc.br}

\affiliation{Departamento de F\'isica, CFM - Universidade Federal de Santa Catarina; \\ C.P. 476, CEP 88.040-900, Florian\'opolis, SC, Brazil.}

\author{Zeinab Rezaei \orcidlink{0000-0002-0646-0425}}
\email{zrezaei@shirazu.ac.ir}

\affiliation{Department of Physics, School of Science, Shiraz University, Shiraz 71454, Iran. }
\affiliation{Biruni Observatory, School of Science, Shiraz University, Shiraz 71454, Iran.}

\author{Tobias Frederico \orcidlink{0000-0002-5497-5490}} 
\email{tobias@ita.br}

\affiliation{Instituto Tecnológico de Aeronáutica,\\ CEP 12.228-900, São José dos Campos, SP, Brazil}

\title{Proto-neutron stars with quark cores} 

\begin{abstract}
This work investigates the evolution of proto-neutron stars (PNSs) from birth as neutrino-rich objects to maturity as cold-catalyzed neutrino-poor objects with nucleonic and non-nucleonic degrees of freedom. The focus is on the star's core where the nucleons, hyperons, and the $\Delta$-isobars are expected to dissolve into a ``soup" of deconfined quarks, at higher baryon densities, to establish a possible hadron-quark phase transition. We separately calculate the nuclear equations of state (EoS) for the hadronic matter (composed of all the baryon octet and $\Delta$-isobars) and the strange quark matter (SQM) under the same thermodynamic conditions characteristic of PNS and proto-strange star (PSS) evolution and construct the hybrid EoS using Maxwell's construction. The study allows us to determine the hadron-quark phase transitions along the evolution lines of the star. We observed a phase transition from hadronic matter to quark matter (QM) phase when the neutrinos have completely escaped from the star's core. The EoSs utilized are constrained to meet the $2\,\rm M_\odot$ threshold in accordance with the observational data.
\end{abstract}

\maketitle

\section{Introduction}
Core-collapse supernova is a suitable laboratory for probing the emission of light particles with masses within $\lesssim 100\,\rm MeV$ \cite{RAFFELT19901, Raffelt:1987yt}. For instance, the observation of neutrino burst in SN 1987A event \cite{Kamiokande-II:1987idp, Hirata:1988ad, PhysRevLett.58.1494} is a clear case in point. This observation set a milestone in particle physics leading to a further constraint on the properties of neutrinos \cite{Kolb:1987qy, Raffelt:1990yu}. Further analysis of the data led to constraints on particles such as axions \cite{Turner:1987by, Chang:2018rso, Lucente:2022vuo}, muonic bosons \cite{Caputo:2021rux}, dark photons \cite{DeRocco:2019njg}, gravitons \cite{Hannestad:2001jv}, unparticles \cite{Hannestad:2007ys}, scalars mixed with Higgs boson \cite{Balaji:2022noj}, and particles escaping from extra dimension \cite{Friedland:2007yj}, largely because, the emission of these particles are suspected to affect the time duration of the neutrino burst. Comparatively, the most explored among these particles is the quantum chromodynamic (QCD) axions otherwise known as the axion-like-particles in the literature \cite{Bharucha:2022lty, Lella:2022uwi}.
{Additionally, the emergence of data from the first authoritative neutron star (NS) binary merger led to the observation of gravitational wave in the event GW170817 \cite{LIGOScientific:2017vwq, LIGOScientific:2018cki}, which with the additional data from the Neutron Star Interior Composition Explorer (NICER) observatory \cite{Riley:2019yda, Miller:2019cac, Riley:2021pdl, Miller:2021qha} have lead to strong constraints on cold EoS that governs the structure of NSs.} These constraints clearly define the mass threshold within which neutron stars can be explored and vital boundaries on the NS radii. {Also, they have ruled out many phenomenological models for  EoSs (see e.g. \cite{Akmal:1998cf}) that do not satisfy these new constraints.} Consequently, the focus has been directed to the EoSs that satisfy these constraints. These EoSs are being confronted with the current observational data to narrow down to more realistic EoSs for multipurpose applications in the future (See a review of relativistic mean field models in \cite{Menezes:2021jmw}, and other forms of EoSs in \cite{CompOSECoreTeam:2022ddl}).  

The main ingredients for exploring high energy astrophysical phenomena such as core-collapse supernovae \cite{Burrows:2012ew, Janka:2016fox}, PNSs \cite{Prakash:1996xs, Janka:2006fh} and binary compact star mergers involving at least one NS \cite{Baiotti:2016qnr, Shibata:2019wef, Radice:2020ddv} are relationships between pressure ($P$), energy density ($\varepsilon$), baryon density ($n_B$) and temperature ($T$). The connection between these quantities and the knowledge of matter composition gives rise to the EoS used in exploring PNSs \cite{Lattimer:2015nhk, Oertel:2016bki}. There is a growing interest in recent years in these astrophysical phenomena motivating several studies intended to provide a precise description of matter under extreme conditions of temperature and density. Owing to the high temperature and density variations in the stellar core during the supernova explosion and binary merger remnants \cite{Fischer:2013eka, Perego:2019adq}, the determination of their EoSs requires theoretical and numerical efforts. 

Another phenomenon of interest is the appearance of non-nucleonic degrees of freedom in the core of compact objects. The leading non-nucleonic degree of freedom theoretically favored to be present in the stellar core has long been conjectured to be the quark matter \cite{Blaschke:2018mqw, Annala:2019puf}. The major doubt remains to be established, at what stage does the quark core begin to appear in the stellar matter during its evolution? This is one of the many questions we will attempt to answer with this contribution. In \cite{PhysRevC.100.015803}, the authors attempt to answer a similar question using a different model while, in \cite{Shao:2011nu}, the authors undertake a similar study considering nucleons and the Gibbs construction of hybrid stars, which differs from the approach we intend to use in this work.  However, in \cite{PhysRevC.100.015803}, they concluded that there are no free quarks in hot stars within their model framework. The aim is to go beyond the usual zero-temperature hybrid neutron stars that have been widely investigated in the literature \cite{Issifu:2023ovi, Karimi:2022kcx, Cierniak:2021knt} to include thermodynamic conditions relevant for hot QM \cite{Issifu:2023qoo} phase transition during the stellar evolution.

The deconfinement phase of quarks and gluons is believed to exist at the two extreme ends on the QCD phase diagram depending on the magnitude of the temperature and the chemical potential. A smooth crossover is observed in lattice QCD calculations at high temperature and low chemical potential ($T\gg \mu$) giving rise to a phase transition from a nuclear matter to quark matter phase \cite{Philipsen:2012nu, Borsanyi:2013bia, Bellwied:2015rza}. On the other hand, as the temperature decreases and the chemical potential increases ($T\ll \mu$) another phase transition (perhaps of first order \cite{Fukushima:2010bq, Most:2018eaw}) from hadronic to quark matter is observed. The former is likely to be the case in the early universe and the latter is the phenomenon believed to occur in the inner core of (proto-) neutron stars \cite{Roark:2018uls} or merger remnants. The region between these two extreme ends can be explored through relativistic heavy-ion collision experiments.

The (proto-) neutron stars are mostly composed of protons and neutrons at low temperatures and {high baryon densities}. However, heavy and strangeness-rich baryons such as the hyperons and even nonstrange $\Delta$-isobars are theorized to appear deep towards the inner core \cite{Marquez:2022gmu}. 
As the baryon density keeps increasing and the particles get tightly parked, their dissociation into a deconfined quark matter becomes energetically favorable. This suggests the possibility of different phases of matter inside the NS: hadronic matter at low baryon densities and quark matter phase at higher baryon densities \cite{Glendenning:1992vb, Annala:2019puf}.

In this work, we investigate the possibility of a hadron-quark phase transition in superdense matter during the evolution of NS from birth to maturity at zero temperature. The main goal is to establish the regime at which the deconfined quark matter begins to appear in the stellar matter during the star's evolution. We broadly considered two regimes: the neutrino-trapped regime and the neutrino-transparent one after all the neutrinos have escaped from the stellar core. The quark matter was described with the density-dependent quark model (DDQM) \cite{Backes:2020fyw, Wen:2005uf} using the free model parameters determined through Bayesian inference \cite{daSilva:2023okq}, this model was applied in studying PSSs in \cite{Issifu:2023qoo}. On the other hand, the hadronic matter is described by the relativistic mean-field approximation model using the density-dependent couplings adjusted by the density-dependent meson-nucleon coupling parameterization \cite{PhysRevC.71.024312, Roca-Maza:2011alv}. This model has also been widely used in studying PNSs in the literature (see e.g. \cite{PhysRevC.100.015803, Issifu:2023qyi, Raduta:2020fdn} and references therein). In particular, we adopt the EoS calculated in \cite{Issifu:2023qyi} to aid our current study. The two models were studied under similar thermodynamic conditions characteristic of PNS (PSS) evolution \cite{Shao:2011nu, Prakash:1996xs} and the hadron-quark phase transition investigated through Maxwell's construction of hybrid stars \cite{Han:2019bub}. This approach predicts a sharp phase transition from hadronic to quark matter phase when chemical and mechanical equilibrium are established between both matter phases.

{The hadronic matter phase is presumed to be composed of nucleons, hyperons, and $\Delta$-isobars. Their corresponding equations of motion are solved using an improved and extended parameterization of the baryon-meson couplings with density-dependent constraints \cite{Lopes:2022vjx}. The quark matter, on the other hand, is composed of up ($u$), down ($d$), and strange ($s$) quarks. The baryon density at which the SQM starts appearing in the stellar matter is also assumed to be several times greater than the nuclear saturation density ($n_0=0.152\,\rm fm^{-3}$). Moreover, our EoSs satisfy the $2\,\rm M_\odot$ threshold of NSs imposed by the discovery of  PSR J0740+6620 \cite{NANOGrav:2019jur, Fonseca:2021wxt},  PSR J2215+5135 \cite{Linares:2018ppq}, and PSR J0952-0607 \cite{Romani:2022jhd} pulsars and the detection of the gravitational wave in event GW170817 by LIGO/Virgo Collaborations \cite{LIGOScientific:2017vwq, LIGOScientific:2016aoc, LIGOScientific:2018cki, LIGOScientific:2018hze}.

The paper is organized as follows, in Sec.~\ref{m} we present the model that forms the basis of the investigations. The section is divided into various subsections to thoroughly discuss each model and its properties. In Subsec.~\ref{hm}, we describe the hadronic model and discuss it in detail, in Subsec.~\ref{qm}, we present the general overview of the quark model and separately discuss the zero-temperature formalism of the model in \ref{ztf} and the finite temperature formalism in \ref{ftf}. In Subsec.~\ref{psm}, we present the properties of the stellar matter used in the analysis, and in Subsec.~\ref{chs}, we discuss the procedure used in constructing the hybrid stars and how to determine the phase transitions. We present our results in Sec.~\ref{ra} and the final remarks in Sec.~\ref{fr}. } 

\section{The models}\label{m}
This study adopts separate, hadronic, and quark matter models used to calculate EoS under thermodynamic conditions characteristic of PNSs and PSSs evolution, respectively. We use the relativistic model within the mean-field approximation to describe the hadronic matter with density-dependent couplings adjusted by the density-dependent meson-nucleon coupling (DDME2) parameterization \cite{PhysRevC.71.024312, Roca-Maza:2011alv}. On the other hand, we use the DDQM \cite{daSilva:2023okq, Peng:1999gh, Xia:2014zaa} to describe the quark matter. We calculate the EoSs from these two models differently and then construct the hybrid EoSs by establishing chemical and mechanical equilibrium between the hadronic and the quark matter phases which serves as a condition for determining phase transition (this will be elaborated in subsequent sections) between the two matter phases. 
We will briefly discuss the hadronic and quark models separately below, before discussing how the hybrid EoS is constructed from these two models in the subsequent sections.

\subsection{The hadronic model}\label{hm}

We use the relativistic model within the mean field approximation to study the hadronic matter comprising the baryon octet, $\Delta$-isobars, leptons, and mesons that mediate the strong interactions among the hadrons in a Lagrangian density formalism
\begin{equation}
     \mathcal{L}_{\rm RMF}= \mathcal{L}_{H}+ \mathcal{L}_{\Delta}+ \mathcal{L}_{\rm m}+ \mathcal{L}_{\rm L},
\end{equation}
where the relativistic mean field (RMF) Lagrangian, $\mathcal{L}_{\rm RMF}$ consists of the baryon octet, $\mathcal{L}_H$, $\Delta$-isobars, $\mathcal{L}_{\Delta}$, mesons, $\mathcal{L}_{\rm m}$, and leptons, $\mathcal{L}_{\rm L}$. A Dirac-like Lagrangian is used to describe the spin-1/2 baryon octet in the form 
\begin{align}\label{a}
 \mathcal{L}_{H}= {}& \sum_{b\in H}  \bar \psi_b \Big[  i \gamma^\mu\partial_\mu - \gamma^0  \big(g_{\omega b} \omega_0  +  g_{\phi b} \phi_0+ g_{\rho b} I_{3b} \rho_{03}  \big)\nonumber \\
 &- \Big( m_b- g_{\sigma b} \sigma_0 \Big)  \Big] \psi_b,
\end{align}
where $\sigma$ is a scalar meson, $\omega$ and $\phi$ are vector mesons (with $\phi$ carrying a hidden strangeness), they are both isoscalars and $\Vec{\rho}$ is a vector-isovector meson. The subscript `0' in the meson fields in $\mathcal{L}_{H}$ indicates their mean-field version. On the other hand, the Rarita-Schwinger--type Lagrangian is used to describe the spin-3/2 baryon decuplet in the form 
\begin{align}\label{a1}
        \mathcal{L}_{\Delta}={}& \sum_{d\in \Delta}\Bar{\psi}_{d}\Big[i \gamma^\mu \partial_\mu- \gamma^0\left(g_{\omega d}\omega_0 + g_{\rho d} I_{3d} \rho_{03} \right) \nonumber\\&-\left(m_d-g_{\sigma d}\sigma_0 \right)\Big]\psi_{d},
\end{align}
due to the additional vector-valued spinor component relative to the spin-1/2 Dirac spinors. That notwithstanding, it has been demonstrated in \cite{dePaoli:2012eq} that the spin-3/2 model equations of motion can be written in the same way as the spin-1/2 particles in the regime of the RMF approximation. Additionally, the mesonic fields that mediate the strong interactions between the hadronic fields are described by the mean-field Lagrangian density
\begin{align}\label{a11}
 \mathcal{L}_{\rm m}= - \frac{1}{2} m_\sigma^2 \sigma_0^2  +\frac{1}{2} m_\omega^2 \omega_0^2  +\frac{1}{2} m_\phi^2 \phi_0^2 +\frac{1}{2} m_\rho^2 \rho_{03}^2,
\end{align}
in which, the $m_i$ are the meson masses tabulated in Tab.~\ref{T}. Finally, the free leptons are introduced through the Dirac Lagrangian density for free particles
\begin{equation}\label{l1}
    \mathcal{L}_{\rm L} = \sum_L\Bar{\psi}_L\left(i\gamma^\mu\partial_\mu-m_L\right)\psi_L,
\end{equation}
where the summation index $L$ runs over all the leptons present. For a cold catalyzed stellar matter, we consider only electrons ($e$) and muons ($\mu$) i.e. $L\in(e,\,\mu)$ with a degeneracy of $\gamma_L = 2$. In the case of fixed entropy stellar matter, we consider the electron neutrino $\nu_e$ in addition, with $\gamma_L =1$, neglecting muons (when $\nu_e$'s are trapped in the stellar matter) and their corresponding neutrinos since muons only become relevant in the stellar matter when all the electron neutrinos have escaped from the core of the star consistent with supernova physics~\cite{Malfatti:2019tpg, Prakash:1996xs}.

Following the DDME2 parameterization \cite{PhysRevC.71.024312}, the meson-baryon couplings are adjusted by the expression
\begin{equation}
    g_{i b} (n_B) = g_{ib} (n_0)a_i  \frac{1+b_i (\eta + d_i)^2}{1 +c_i (\eta + d_i)^2},
\end{equation}
where $i=\sigma, \omega, \phi$ and 
\begin{equation}
    g_{\rho b} (n_B) = g_{\rho b} (n_0) \exp\left[ - a_\rho \big( \eta -1 \big) \right],
\end{equation}
with $\eta =n_B/n_0$. 
The model parameters are determined by fitting to the experimental data of bulk nuclear properties at the saturation density ($n_0=0.152\,{\rm fm^{-3}}$). The calculated binding energy ($E_B = -16.4\,{\rm MeV}$), the compressibility modulus, ($K_0 = 251.9\,{\rm MeV}$), symmetry energy ($J = 32.3\,{\rm MeV}$) and its slope ($L_0 = 51.3 \,{\rm MeV}$) also agree fairly with the recent constraints on bulk symmetric nuclear matter properties \cite{Dutra:2014qga, Reed:2021nqk, Lattimer:2023rpe}.

As the free parameters of the model are determined through fitting to pure nucleonic matter, the extension to determine the meson-baryon coupling is done relative to the nucleon couplings through the ratio $\chi_{ib} = g_{ib}/g_{iN}$ (the couplings for the nucleons at saturation density have been tabulated in Tab.~\ref{T}). There are several forms of calculating these couplings using either SU(3) or SU(6) flavor arguments which are extensively documented in the literature (see e.g. \cite{Glendenning:1991es, Stone:2019blq, Lopes:2020rqn, RevModPhys.38.215, Weissenborn:2011kb, DOVER1984171, Schaffner:1993qj}). However, we adopt the results presented in \cite{Lopes:2022vjx} where the authors have calculated the meson-hyperon couplings and extended it to the meson-$\Delta$-isobar couplings in a model-independent manner (see Tab.~\ref{T1}). The particle potentials used in fitting these couplings are $U_\Lambda = -28\, {\rm MeV}$, $U_\Sigma= 30$~MeV, $U_\Xi=-4$~MeV, and $U_\Delta\approx -98$~MeV. The couplings displayed on Tab.~\ref{T1} correspond to $\alpha_v = 0.5$, {the only free parameter of SU(2) and SU(3) groups used in fixing the baryon-meson couplings as presented in Tab.~II } 
of \cite{Lopes:2022vjx}  which has been recalculated taking into account the isospin projections in  Eqs.~(\ref{a}) and (\ref{a1}). 

The relevant thermodynamic quantities required to calculate the EoS can be derived from the Lagrangian densities. The baryon density can be determined as 
\begin{equation}\label{a2}
n_b = \gamma_b \int \frac{d^3 k}{(2\pi)^3}  \left[f_{b\,+}(k) - f_{b\,-}(k)  \right]
\end{equation}
with $\gamma_b = 2$, the degeneracy of the spin-1/2 baryon octet while $f_{\rm b \pm}(k)$ is the Fermi--Dirac distribution function given by 
\begin{equation}
    f_{b \pm}(k) = \frac{1}{1+\exp[(E_b \mp \mu^\ast_b)/T]} \nonumber
\end{equation}
where $E_b = \sqrt{k^2+m_b^{*2}}$ is the single particle energy with effective baryon mass $m_b^*$ and an effective chemical potential $\mu_b^*$. We can retrieve the relation for the $\Delta$-isobars by interchanging $b\leftrightarrow d$, here, the $E_d = \sqrt{k^2+m_d^{*2}}$ with $\gamma_d = 4$. The effective chemical potentials can be expressed as 
\begin{align}
    \mu_{b,d}^\ast &= \mu_{b,d}- g_{\omega {b,d}} \omega_0 - g_{\rho {b,d}} I_{3{b,d}} \rho_{03} - g_{\phi {b}} \phi_0 - \Sigma^r,
\end{align}
where $\Sigma^r$ is the rearrangement term arising from the thermodynamic consistency of the model. Its explicit form reads:
\begin{align}
    \Sigma^r ={}& \sum_b \Bigg[ \frac{\partial g_{\omega b}}{\partial n_b} \omega_0 n_b + \frac{\partial g_{\rho b}}{\partial n_b} \rho_{03} I_{3b}  n_b+ \frac{\partial g_{\phi b}}{\partial n_b} \phi_0 n_b \nonumber \\
    &- \frac{\partial g_{\sigma b}}{\partial n_b} \sigma_0 n_b^s + b\leftrightarrow d\Bigg].
\end{align}
The effective masses are given by 
\begin{equation}\label{a3}
    m_b^\ast =m_b- g_{\sigma b} \sigma_0, \quad\quad m_d^\ast =m_d- g_{\sigma d} \sigma_0,
\end{equation}
with a scalar density
\begin{equation}
    n_{b}^s =\gamma_b \int \frac{d^3 k}{(2\pi)^3} \frac{m^\ast_b}{E_b} \left[f_{b\,+}(k) + f_{b\,-}(k)  \right],
\end{equation}
similar expressions can be obtained for the $\Delta$-isobars by interchanging $b$ and $d$ with the appropriate degeneracy. It is important to note that the expressions in Eqs. (\ref{a2}) and (\ref{a3}) can be used to calculate the densities of the leptons taking into consideration that the leptons have constant mass and chemical potentials with $\gamma_L$ degeneracy. The mean-field equations of motion for Eqs.~(\ref{a}), (\ref{a1}) and (\ref{a11}) are 
\begin{align}
    &m^2_\sigma \sigma_0 = \sum_b g_{\sigma b}n^s_b + \sum_d g_{\sigma d}n_d^s,\\
    &m^2_\omega\omega_0 = \sum_b g_{\omega b}n_b + \sum_d g_{\omega d}n_d,\\
    &m^2_\phi \phi_0 = \sum_b g_{\phi b}n_b, \\
    &m^2_\rho\rho_{03} = \sum_b g_{\rho b}n_bI_{3b} + \sum_d g_{\rho d}n_dI_{3d}.
\end{align}
The total energy density and the pressure of the system can be determined through 
\begin{flalign}\label{1a}
    \varepsilon_B&=  \varepsilon_b + \varepsilon_m + \varepsilon_d +\varepsilon_L,\\
    P_B&=  P_b + P_m +P_d + P_L + P_r\label{1b},
\end{flalign}
where the subscripts denote matter under consideration. The kinetic contributions from the baryon octet are
 \begin{equation}\label{eq:ener_b}
    \varepsilon_b= \gamma_b \int \frac{d^3 k}{( 2\pi)^3} E_b \left [f_{b+}(k) +f_{b-}(k) \right],
\end{equation}

\begin{equation}\label{eq:press_b}
    P_b= \gamma_b \int \frac{d^3 k}{( 2\pi)^3} \frac{k^2}{E_b} \left [ f_{b+}(k) +f_{b-}(k) \right],
\end{equation}
a similar expression can be used for $\varepsilon_d$, $\varepsilon_L$, $P_d$ and $P_L$ with the exchange of $b$ with $d$ and $L$ with the appropriate degeneracies. The contribution from the meson fields can be calculated from the energy-momentum tensor, leading to 
\begin{equation}\label{eq:ener_m}
    \varepsilon_m=  \frac{m_\sigma^2}{2} \sigma_0^2+\frac{m_\omega^2}{2} \omega_0^2 +\frac{m_\phi^2}{2} \phi_0^2  + \frac{m_\rho^2}{2} \rho_{03}^2,
\end{equation}
and 
\begin{equation}\label{eq:press_m}
    P_m= -  \frac{m_\sigma^2}{2} \sigma_0^2 +\frac{m_\omega^2}{2} \omega_0^2 +\frac{m_\phi^2}{2} \phi_0^2  + \frac{m_\rho^2}{2} \rho_{03}^2.
\end{equation}
The total pressure received a correction
\begin{equation}
    P_r = n_B\Sigma^r,
\end{equation}
known as the rearrangement self-energy term which occurs as a consequence of thermodynamic consistency, which can be verified directly from
\begin{equation}
    P_B = n_B^2\dfrac{\partial }{\partial n_B}\left( \dfrac{\varepsilon_B}{n_B}\right).
\end{equation}
From the above thermodynamic quantities, we can determine the free energy expression $\mathcal{F}_B= \varepsilon_B - Ts_B$, where $T$ is temperature and $s_B=S/n_B$ is the entropy per baryon given by 
{\begin{align}
     s_B = \frac{\varepsilon_B +P_B- \sum_{b} \mu_{b} n_{b} -\sum_{d} \mu_{d} n_{d} -\sum_{L} \mu_{L} n_{L} 
}{T}.
\end{align}
Simplifying the above expressions using $\beta$-equilibrium and charge neutrality conditions yields;
\begin{equation}
    s_BT=P_B+\varepsilon_B -n_B\mu_B,
\end{equation}
for neutrino-transparent stellar matter, where $\mu_B$ is the baryon chemical potential and 
\begin{equation}
    s_BT=P_B+\varepsilon_B -n_B\mu_B-\mu_{\nu_e}(n_{\nu_e} +n_e),
\end{equation}
for neutrino-trapped matter, with $\mu_{\nu_e}$ the neutrino chemical potential, $n_{\nu_e}$ and $n_e$ are the neutrino and the electron number densities. 
}
\begin{table}
\begin{center}
\begin{tabular}{ |c| c| c| c| c| c| c| }
\hline
 meson($i$) & $m_i(\text{MeV})$ & $a_i$ & $b_i$ & $c_i$ & $d_i$ & $g_{i N} (n_0)$\\
 \hline
 $\sigma$ & 550.1238 & 1.3881 & 1.0943 & 1.7057 & 0.4421 & 10.5396 \\  
 $\omega$ & 783 & 1.3892 & 0.9240 & 1.4620 & 0.4775 & 13.0189  \\
 $\rho$ & 763 & 0.5647 & --- & --- & --- & 7.3672 \\
 \hline
\end{tabular}
\caption {DDME2 parameters.}
\label{T}
\end{center}
\end{table}

\begin{table}
\begin{center}
\begin{tabular}{ |c | c| c| c| c| } 
\hline
 b,d & $\chi_{\omega b,d}$ & $\chi_{\sigma b,d}$ & $\chi_{\rho b,d}$ & $\chi_{\phi b}$  \\
 \hline
 $\Lambda$ & 0.714 & 0.650 & 0 & -0.808  \\  
$\Sigma^0$ & 1 & 0.735 & 0 & -0.404  \\
  $\Sigma^{-}$, $\Sigma^{+}$ & 1 & 0.735 & 0.5 & -0.404  \\
$\Xi^-$, $\Xi^0$  & 0.571 & 0.476 & 0 & -0.606 \\
  $\Delta^-$, $\Delta^0$, $\Delta^+$, $\Delta^{++}$   & 1.285 & 1.283 & 1 & 0  \\
  \hline
\end{tabular}
\caption {The ratio of the baryon coupling to the corresponding nucleon coupling for hyperons and $\Delta$s.}
\label{T1}
\end{center}
\end{table}

\subsection{The quark Model}\label{qm}
The DDQM model is constructed through the QCD Hamiltonian density 
\begin{equation}\label{b}
    H_{\rm QCD} = H_k + \sum_{i = u,\,d,\,s}m_{i0}\bar{q}q + H_I,
\end{equation}
where $H_k$ is the kinetic term, $q(\bar{q})$ are the quark fields, and $m_{i0}$ is the bare quark mass together with the equivalent Hamiltonian density 
\begin{equation}\label{b1}
    H_{\rm eqv} = H_k + \sum_{i = u,\,d,\,s}m_{i}\bar{q}q.
\end{equation}
To maintain the constant energy of the free quark system, the equivalent quark mass ($m_i$) term must take the form
\begin{equation}
    m_i = m_{i0}+m_I,
\end{equation}
with $m_I$ being associated with the interaction term. To determine $m_i$, we ensure that both $H_{\rm QCD}$ and $H_{\rm eqv}$ reproduce the same eigenvalues for the same eigenstates,  $|n_B\rangle$, therefore 


\begin{equation}
    \langle n_B|H_{\rm QCD}|n_B \rangle = \langle n_B|H_{\rm eqv}|n_B \rangle,
\end{equation}
must be satisfied. Additionally, taking the vacuum expectation values, $|0\rangle$, of both expressions and taking the difference, yields
\begin{equation}\label{vf}
    m_i = m_{i0} + \dfrac{\langle H_I \rangle_{n_B} - \langle H_I \rangle_0}{\sum_{q}\Big[\langle \bar{q}q \rangle_{n_B} - \langle \bar{q}q \rangle_0\Big]} ={ m_{i0}+m_I},
\end{equation}
with $\langle n_B|H_I|n_B \rangle = \langle H_I \rangle_{n_B}$, $\langle n_B|q\bar{q}|n_B \rangle = \langle \bar{q}q \rangle_{n_B}$ and the vacuum expectation values are expressed in the same way with subscript `0'. The second term in Eq.(\ref{vf}) is precisely the $m_I$ term. Since $m_I$ depends on the interaction energy density and the quark condensate, it is flavor independent. The interaction energy density can be expressed in terms of the average inter-particle confining potential and the baryon density as 
\begin{equation}\label{v2}
    \langle H_I \rangle_{n_B} - \langle H_I \rangle_0 = 3n_BV_c(r_*),
\end{equation}
where factor 3, represents the three quark flavors present, $V_c(r_*)$ is the confining potential and $r_*$ is the average inter-particle separation distance. Also, from the model-independent quark condensate in nuclear matter \cite{PhysRevLett.67.961, PhysRevC.45.1881}, we can express the denominator as
\begin{equation}\label{v3}
    \dfrac{\langle\bar{q}q\rangle_{n_B}}{\langle\bar{q}q\rangle_0} = 1-\dfrac{n_B}{n_B^*},
\end{equation}
where $n_B^*$ is a constant associated with the pion mass $M_\pi$, its decay constant $f_\pi$, and the pion-nucleon sigma term $\sigma_N$ in a form,
\begin{equation}
    n^*_B=\dfrac{M_\pi^2f_\pi^2}{\sigma_N}.
\end{equation}
Therefore, the general structure of (\ref{vf}) becomes 
\begin{equation}\label{d1}
    m_i = m_{i0} + \dfrac{D}{(n_B)^z},
\end{equation}
with
\begin{equation}\label{vf1}
    D\sim - \dfrac{ 3\sigma_0 n_B^*}{ \sum_q\langle\bar{q}q\rangle_0}, 
\end{equation}
where $\sigma_0$ is the proportionality constant of the confining potential representing the string tension. The last term in (\ref{d1}) is precisely the $m_I$ satisfying the condition
\begin{equation}\label{va}
\lim_{n_B \rightarrow 0} m_I \rightarrow \infty;\qquad{\text{color confinement condition}}.
\end{equation}
{In addition, the $M_\pi$, $f_\pi$, $\sigma_N$, $\sigma_0$ and the constant quark condensate $\sum_q\langle\bar{q}q\rangle_0$ are all absorbed into the constant $D$ whose value is determined by fitting the astrophysical observation data (see details of how it was fitted through Bayesian inference in \cite{daSilva:2023okq})}. The potential $V_c(r_*)=\sigma r^n$ also satisfies the same conditions considering that $r_*^n\sim 1/n_B^z$, where $n$ and $z$ are arbitrary constants determined by dimensional analysis. To obtain quark stars with masses comparable to the observed neutron star (NS) mass threshold \cite{NANOGrav:2019jur, Fonseca:2021wxt}, we employ the famous Cornell potential \cite{PhysRevLett.34.369, Bali:2000gf}, given by { $V_c(r_*)=-(\beta/r_*)+\,\sigma r_*$}, which incorporates both the confinement and the asymptotic freedom behavior determined from QCD lattice calculations. Here, an additional term is included in (\ref{d1}), where $\sigma$ represents the string tension and $\beta$ is a constant representing the deconfinement strength. Hence, the new expression becomes 
\begin{equation}\label{d}
    m_i = m_{i0} + \dfrac{D}{(n_B)^z} + C(n_B)^z,
\end{equation}
the additional term will contribute to repulsive pressure;
\begin{equation}
    P_i \sim (n_B)^2\dfrac{\partial m_i(n_B)}{\partial n_B},
\end{equation}
to balance the attractive pressure generated by the confining term to hold the star from collapsing as it reaches its maximum possible mass. Hereafter, we shall choose $z=1/3$ in conformity with the lattice QCD (LQCD) confining potential, $V_c(r_*) = \sigma r_*$ as well as the Cornell potential for confining heavy quarks, here $n=1$ as well. That notwithstanding, this cubic root mass scaling law has been widely applied in studying quark stars using the DDQM by several authors (see e.g. \cite{Peng:1999gh, Xia:2014zaa, daSilva:2023okq} and references therein).


To introduce temperature into the system, we follow the LQCD approach, where the temperature is introduced through the string tension \cite{Bicudo:2010hg, Kaczmarek:1999mm, Digal:2003jc}. In this case, the string tension decreases with increasing temperature and breaks at the critical temperature ($T_C$) where the particles become asymptotically free. Thus, the normalized string tension takes the form 
\begin{equation}
    \dfrac{\sigma(T)}{\sigma_0} = \left[1-\dfrac{T^2}{T_C^2}\right]^\alpha, 
\end{equation}
where $\alpha$ is an arbitrary constant whose value depends on the model under consideration. At $T=0$, we recover the $\sigma(T=0) = \sigma_0$ and it vanishes at $T=T_C$. For the sake of the current work, we will adopt $\alpha =1/2$ (see e.g. \cite{Philipsen:2007rj, Ilgenfritz:2017kkp, Bicudo:2010hg} and references therein). Therefore, the full expression for the equivalent mass term becomes 
\begin{equation}
      m_i = m_{i0} + \dfrac{D}{n_B^{1/3}}\left[1-\dfrac{T^2}{T_C^2}\right]^{\frac12}  + Cn_B^{1/3}\left[1-\dfrac{T^2}{T_C^2}\right]^{-\frac12},
\end{equation}

similar to the mechanism introduced in Refs.~\cite{Karsch:2000kv, Laermann:2003cv}. This approach has been adopted by several authors (see e.g. \cite{Wen:2005uf, Chen:2021fdj, Chen:2023rza, Chu:2017huf, Issifu:2023qoo, Issifu:2024zvq}) to introduce temperature into the equivalent mass model arguing that $D$ (confining term) decreases with temperature, on the other hand, $C$ (the asymptotic freedom term) increases with temperature to stabilize the mass of the star. The quark masses adopted for our calculations are $m_u = 2.16\,\rm MeV$, $m_d = 4.67\,\rm MeV$, and $m_s = 93.4\,\rm MeV$, as documented in \cite{ReviewOfParticlePhysics}. The free model parameters for the quark model are $\sqrt{D}=128.4\,\rm MeV$, and $C=0.8$. These values represent a slight adjustment {around the stability window} from the findings reported in \cite{daSilva:2023okq}, aimed at rendering quark stars marginally unstable, preventing the rapid transformation of hybrid stars into quark stars. {Specifically, we modified the value of \( \sqrt{D} \) from 127.4 MeV to 128.4 MeV to introduce the slight instability mentioned above.
In contrast, the model parameters determined in \cite{daSilva:2023okq} were based on absolute stable strange quark matter using the Bodmer-Witten conjecture \cite{PhysRevD.4.1601, Witten:1984rs}, thus capable of transforming a hybrid star into a strange quark star.} The next step is to check for the thermodynamic consistency of the mass formulas derived above.

\subsubsection{Zero Temperature Formalism}\label{ztf}
To ensure that the model is thermodynamically consistent at zero temperature, we define the thermodynamic free energy ($F$) where the free particle mass is replaced with $m_i(n_B)$ and the real chemical potential $\mu_i$ replaced with an effective chemical potential $\mu^*_i$ therefore,
\begin{equation}
    F = \Omega_0 \left( \{ \mu^*_i \}, \{ m_i \} \right) + \sum_i \mu^*_i n_i,
\end{equation}
where $\Omega_0$ is the thermodynamic potential given explicitly by
\begin{equation}
    \Omega_0 = - \sum_i \frac{\gamma_i}{24 \pi^2} \left[ \mu^*_i \nu_i \left( \nu_i^2 -\frac{3}{2} m_i^2 \right) +\frac{3}{2} m_i^4 \ln{\frac{\mu^*_i +\nu_i}{m_i}}\right], 
\end{equation}
with $\gamma_i=6~ (3~ {\rm colors} \times 2~ {\rm  spins})$ the degeneracies of the quarks. The Fermi momenta in terms of $\mu_i^*$ is given as 
\begin{equation}
    \nu_i = \sqrt{\mu^{*2}_i - m_i^2}.
\end{equation}
It follows that the number density of the quarks can be expressed as 
\begin{equation}\label{1t}
    n_i = \dfrac{\gamma_i}{2\pi^2}\int_0^{\nu_i}p^2dp = \dfrac{\gamma_i\nu_i ^3}{6\pi^2},
\end{equation}
with $p$ the momenta of the quarks, the $\mu_i$ in terms of the $\mu_i^*$ is also given as 
\begin{equation} \label{6t}
    \mu_i = \mu^*_i + \sum_j\dfrac{\partial \Omega_0}{\partial m_j}\dfrac{\partial m_j}{\partial n_i} \equiv \mu_i^* - \mu_I,
\end{equation}
where $\mu_I$ is the interacting chemical potential that links $\mu_i$ and $\mu^*_i$ together. From the above expressions, we can determine the energy density ($\varepsilon$) for the system of quarks and its pressure ($P$) through
\begin{equation}
    \varepsilon = \Omega_0 - \sum_i \mu_i^* \frac{\partial \Omega_0}{\partial \mu_i^*},
\end{equation}
and 
\begin{equation}
    P = -\Omega_0 + \sum_{i,j} \frac{\partial \Omega_0}{\partial m_j} n_i \frac{\partial m_j}{\partial n_i}.
\end{equation}
respectively. The interested reader can check Refs~\cite{daSilva:2023okq, Backes:2020fyw, Peng:1999gh} for more detailed derivations.

\subsubsection{Finite Temperature Formalism}\label{ftf}

At a finite temperature, the system of quark matter (QM) depends on the $T,\, n_i$ and the volume ($V$) of the system, consequently, the thermodynamic free energy of the system becomes 
\begin{align}\label{1b1}
    F &= F(T,V,\{n_i\},\{m_i\})\nonumber\\
    &=\Omega_0(T, V,\{\mu^*_i\},\{m_i\}) + \sum_{i=u,d,s}\mu^*_in_i.
    \end{align}
In this expression, the $\Omega_0$ is connected to the independent state variables through 
\begin{equation}\label{1ab}
    n_i = -\dfrac{\partial}{\partial\mu^*_i}\Omega_0(T, V,\{\mu^*_i\},\{m_i\}).
\end{equation}
The explicit expression for the $\Omega_0$ and $n_i$ in terms of particle and anti-particle contributions derived from (\ref{1b1}), are 
\begin{align}
    \Omega^\pm_0 &= - \sum_i\dfrac{\gamma_i T}{2\pi^2}\int_0^\infty p^2dp\bigg[\ln[1+e^{-(\epsilon_i-\mu_i^*)/T}] \nonumber\\
    &+ \ln[1+e^{-(\epsilon_i+\mu_i^*)/T}]\bigg],
\end{align}
and 
\begin{align}
    n_i^{\pm}=-\dfrac{\partial \Omega_0^\pm}{\partial\mu^*_i} &= \sum_i\dfrac{\gamma_i}{2\pi^2}\int_0^\infty p^2dp\bigg[\dfrac{1}{1+e^{(\epsilon_i-\mu_i^*)/T}} \nonumber\\
    &- \dfrac{1}{1+e^{(\epsilon_i+\mu_i^*)/T}}\bigg],
\end{align}
with $\epsilon_i = \sqrt{(p^2+m_i^2)}$, the single particle energy respectively. The $\mu_i$ is expressed in the same way as (\ref{6t}) with its temperature dependence introduced through $\Omega_0(m_i, \mu^*, T)$ and $m_i(n_B, T)$. The other thermodynamic quantities such as the entropy density ($S$),
\begin{equation}\label{4a}
    S_i = -\dfrac{\partial \Omega_0}{\partial T} - \sum_i \dfrac{\partial m_i}{\partial T}\dfrac{\partial\Omega_0}{\partial m_i},
\end{equation}
 the energy density 
 \begin{equation}\label{6b}
    \varepsilon_i = \Omega_0 + \sum_i\mu_i^*n_i - T\dfrac{\partial \Omega_0}{\partial T} - T\sum_i \dfrac{\partial m_i}{\partial T}\dfrac{\partial\Omega_0}{\partial m_i},
\end{equation}
and the pressure 
\begin{align}\label{6a}
    P_i = -\Omega_0 + \sum_{i,j}n_i\dfrac{\partial \Omega_0}{\partial m_j}\dfrac{\partial m_j}{\partial n_i},
\end{align}
can all be derived from (\ref{1b1}) in terms of $T$ and $n_i$. The volume term that appears in (\ref{1b1}) does not affect the pressure considering 
an infinitely large system of quark matter \cite{Wen:2005uf, Peng:2000ff, Xia:2014zaa,  Issifu:2023qoo, Issifu:2024zvq}. {The above expressions are also used to calculate the entropy, pressure, and energy densities of leptons in quark matter. Thus, the total pressure of the quark matter becomes
\begin{equation}
    P = \sum_i P_i + \sum_L P_L,
\end{equation}
the total energy density becomes 
\begin{equation}
    \varepsilon = \sum_i \varepsilon_i + \sum_L \varepsilon_L,
\end{equation}
and the total entropy becomes
\begin{equation}
    S = \sum_i S_i + \sum_L S_L.
\end{equation}
}

\subsection{Stellar Matter Properties }\label{psm}
The hadronic matter comprises nucleons, hyperons, and $\Delta$-isobars, whereas the quark matter consists of $u$, $d$, and $s$ quarks. 
In both scenarios, within the neutrino-trapped region, the prevalent leptons are electrons and their corresponding electron neutrinos ($\nu_e$), as depicted in the initial and intermediate stages of Fig.~\ref{pfs}. The appearance of muons becomes relevant only after the star has transitioned to a neutrino-transparent state by supernova physics \cite{Malfatti:2019tpg, Prakash:1996xs, Weber:2019xvv}. Moreover, the presence of $\tau$-leptons is disregarded due to their substantial mass. We develop numerical codes to independently compute the EoS for both hadronic and quark matter phases and their associated temperature profiles while maintaining fixed values for $S/n_B$ and $Y_{L,e}$. These thermodynamic conditions are selected to accurately describe the evolution of NSs, drawing from relevant literature on the subject \cite{Raduta:2020fdn, Oertel:2016xsn, Sedrakian:2022kgj, PhysRevC.100.015803, Shao:2011nu, Chen:2021edy, Kumari:2021tik}. In both scenarios, we examine $\beta$-equilibrated matter. Consequently, the hadronic matter involves the utilization of the chemical equilibrium conditions:
\begin{align}
    &\mu_\Lambda = \mu_{\Sigma^0} = \mu_{\Xi^0} = \mu_{\Delta^0} = \mu_{n}=\mu_B,\\
    &\mu_{\Sigma^-} = \mu_{\Xi^-} = \mu_{\Delta^-} = \mu_{B}-\mu_Q,\\
    &\mu_{\Sigma^+} = \mu_{\Delta^+} = \mu_p=  \mu_{B}+\mu_Q, \\
    &\mu_{\Delta^{++}}=\mu_{B}+2\mu_Q,
\end{align}
and similar expressions can be found in \cite{Raduta:2021xiz, Raduta:2020fdn, Issifu:2023qyi} where the authors consider hyperons and $\Delta$-isobars in the stellar matter. 
The baryon chemical potential is denoted as $\mu_B$ and the charged chemical potential is defined as $\mu_Q = \mu_p - \mu_n$, where the subscripts correspond to the individual particles present. In the context of neutrino-trapped matter, $\mu_Q$ assumes the following expression
\begin{equation}
  \mu_Q= \mu_{\nu l}-\mu_l , 
\end{equation}
and in the neutrino-transparent regime, it takes on the following form
\begin{equation}
    \mu_Q =-\mu_l,
\end{equation}
where $l$ is the leptons present in that regime. On the other hand, the quark matter was studied following chemical potential expressions 
\begin{equation}
    \mu^*_d = \mu_s^* =\mu_u^*+\mu_e -\mu_{\nu_e}, \quad\text{neutrino trapped matter}\quad
\end{equation}
and
\begin{equation}
    \mu^*_d = \mu_s^* = \mu_u^* + \mu_e, \quad\text{neutrino transparent matter}\quad
\end{equation}
here, the subscripts 
$u,\, d,\, s$ 
represents up, down, and strange quarks. Also, charge neutrality, baryon number conservation, and the lepton number conservation (in the neutrino-trapped region) were separately ensured for both the hadronic and the quark matter phases (see the explicit expressions in Refs.~\cite{Issifu:2023qyi, Issifu:2023qoo}). The lepton numbers are conserved through 
\begin{equation}
    Y_{L,l} = \dfrac{n_l+n_{\nu l}}{n_B},
\end{equation}
where $Y_{L,e} = Y_e + Y_{\nu e}$ and $Y_{L, \mu} = Y_\mu + Y_{\nu_\mu}\approx 0$, following  {the assumptions previously stated based on} supernova physics. The values of $Y_{L, e}$ are primarily influenced by the efficiency of the electron capture reaction occurring during the initial phases of proto-neutron star formation. Additionally, $Y_{L, \mu}\approx 0$ indicates the absence of muons in the stellar matter during the period of neutrino trapping \cite{Prakash:1996xs}. The particle distribution in the stellar matter was computed utilizing
\begin{equation}\label{yi}
    Y_i = \dfrac{n_i}{\sum_in_i},
\end{equation}
where the summation index $i$ runs over all the individual {baryons and the quarks} in each matter phase.

\subsection{Construction of the hybrid stars}\label{chs}

\begin{figure*}[ht!]
  \includegraphics[scale=0.5]{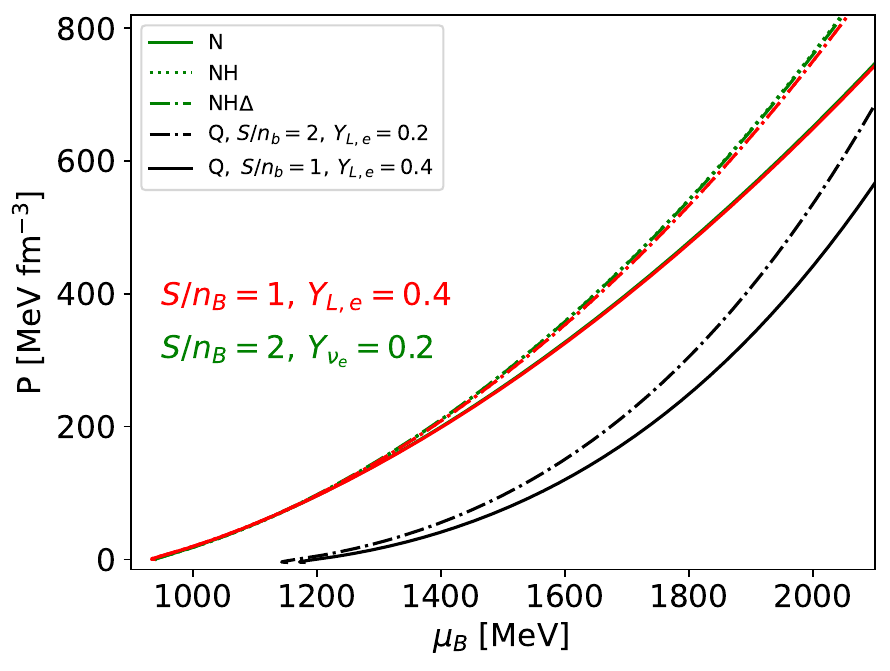}
  \quad
   \includegraphics[scale=0.5]{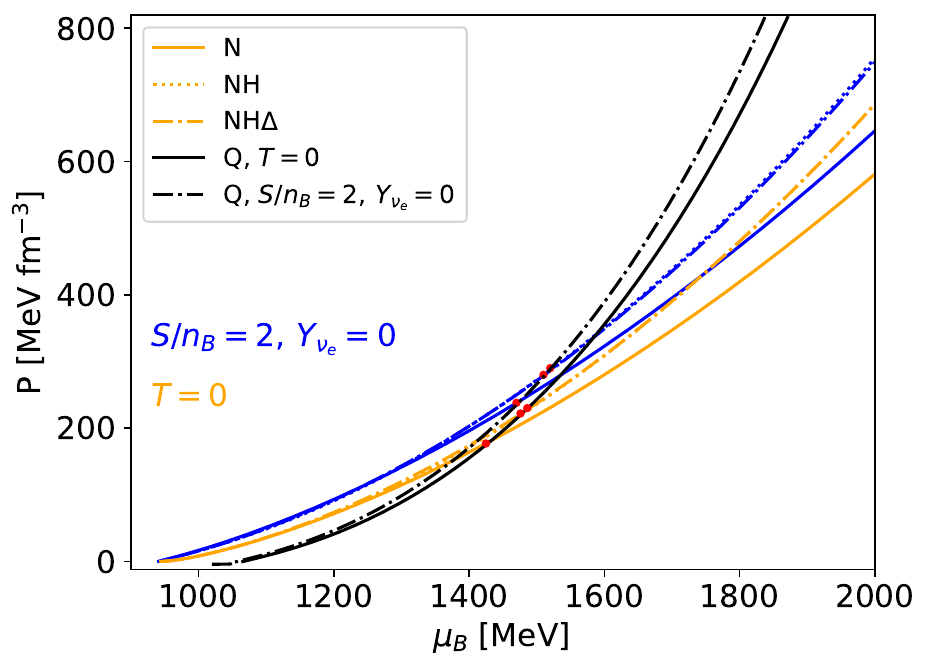}
  
\caption{The graphs depict the pressure as a function of the baryon chemical potential. {We use the colored lines to represent hadronic matter, where solid lines denote nucleons only, dotted lines signify nucleons plus hyperons, and dash-dotted lines represent nucleons plus hyperons plus $\Delta$-isobars and black lines stand for quark matter.} The mechanical and chemical equilibrium points, denoting the phase transition are identified by the intersection between the quark matter and hadronic matter curves. In the left panel, neutrino-rich stellar matter 
curves do not intersect, indicating the absence of a phase transition at this stage. Conversely, in the right panel, where the stellar matter is neutrino-poor, both curves (quark and hadronic matter EoSs) intersect, indicating a phase transition occurring at this stage.
}
    \label{equ}
\end{figure*}

There are generally two ways hybrid stars are constructed in the literature, the Gibbs construction \cite{Shao:2011nu} which involves a mixed phase and a smooth 
transition, and the Maxwell construction which shows a sharp first-order phase transition without a mixed phase -- see Ref.~\cite{Han:2019bub} for an elaborate discussion of the two scenarios. In the microphysical realm, it has been shown that these two approaches exhibit no substantial differences \cite{Maruyama:2007ey, Paoli:2010kc}. In the present investigation, we employ the Maxwell construction, using the two-phase approach wherein the higher density phase is characterized by quark matter. In comparison, the lower density phase consists of hadronic matter comprising baryon octet and $\Delta$-isobars \cite{Issifu:2023qyi, Sedrakian:2022kgj}. In this construction, the EoS for the hadronic phase and the quark matter phase are developed differently and the hybrid star is constructed using chemical and mechanical equilibrium conditions i.e.
\begin{equation}\label{m1}
     P_H=P_q=P_c; \quad\text{mechanical equilibrium}\quad 
\end{equation}
and 
\begin{equation}\label{m2}
    \mu_H=\mu_q=\mu_c; \quad\text{chemical equilibrium}\quad 
\end{equation}
where $P_H$ and $\mu_H$ are the pressure and the chemical potential of the hadronic phase and $P_q$ and $\mu_q$ are the pressure and the chemical potential of the quark phase. {In the case of trapped-neutrinos, the condition for chemical equilibrium is modified such that the Gibbs free energy in both phases can match, thus
\begin{equation}\label{m2}
    \mu_H(b, \nu_e)=\mu_q(i, \nu_e)=\mu_c; \quad\text{neutrino trapped.}\quad
\end{equation}} 
The points of equality between the pressures ($P_c$) and the chemical potentials ($\mu_c$) are referred to as the critical pressure and critical chemical potential respectively (the equilibrium points are shown in Fig.~\ref{equ} with red dots). 
Marking the phase transition point from the hadronic to the QM phase \cite{Issifu:2023ovi, Lopes:2021jpm}. Temperature fluctuations in QM have been established to result in entropies higher than those of hadronic matter due to different particle arrangements and interactions \cite{Ray:2000fa, Sedrakian:2023eqv}. QM has more particle degrees of freedom than hadronic matter, which is effectively composed of confined quarks. Therefore, the fixed entropy in both matter phases has a more pronounced effect on the EoS of QM than on hadronic matter, as illustrated in Fig.~\ref{equ}. Comparing the graphs in both panels, we can infer that the lepton fraction softens the EoS of the QM more significantly than its hadronic counterpart as it impedes the establishment of chemical and mechanical equilibrium between the two matter phases in the left panel.

Even though the existence of hybrid stars was theorized decades ago, the observation of binary NS merger leading to the detection of the gravitational wave in event GW170817 (GW)~\cite{LIGOScientific:2017vwq} and the associated $\gamma$-ray burst, GRB 170817A (GRB)~\cite{LIGOScientific:2017zic, Goldstein:2017mmi} reinforce the belief of its existence. In these instances, the relative time difference between GW and GRB, along with the measured speed of sound of the ejecta from the core of the remnants, offer guidance for categorizing matter phases within NSs. These observations reinforce the theoretical plausibility of a quark core, which could give rise to hybrid NSs. Indeed, the investigation of hybrid stars remains model-dependent, primarily due to the absence of experimental determination or direct observation by any observatory as of the current state. However, despite this limitation, numerous studies exploring the existence of hybrid stars can be found in the literature (see e.g. \cite{Annala:2019puf, Karimi:2022kcx, Cierniak:2021knt, Contrera:2022tqh, Alford:2004pf, Clevinger:2022xzl}).

From the previous sections discussing the hadronic and quark matter models, we compute the EoS for both matter phases and plotted $P_H(\mu_H)$ and $P_q(\mu_q)$ in Fig.\ref{equ} to identify chemical and mechanical equilibrium points, illustrated with red dots in Fig.\ref{equ}. In the {\color{blue}left} panel, we find that the $P_H(\mu_H)$ and $P_q(\mu_q)$ do not cross each other due to the absence of phase transition in this regime when neutrinos are present in the stellar matter, so no equilibrium is established between the two matter phases. 
However, in the right panel, the curves cross each other indicating a phase transition from hadronic to quark matter phase.

\section{Results and Analysis}\label{ra}
\begin{table*}[ht!]
 \footnotesize
\setlength\tabcolsep{1pt}
\begin{center}
\begin{tabular}{|c |c| c| c| c| c| c| c|c|c|c|c|}
 \hline
 \multicolumn{12}{|c|}{Stellar properties} \\
  \hline
  Particle content &S$/$n$_B$ &Y$_{L,e}$ & $M[{\rm M}_{\odot}]$& $M_b[{\rm M}_{\odot}]$ & $R[{\rm km}]$&$\varepsilon_0[{\rm MeV fm^{-3}}]$&$n_c[n_0]$& $T_{co}[{\rm MeV}]$& $T_C[{\rm MeV}]$& $\mu_c[{\rm MeV}]$&$P_c[{\rm MeV fm^{-3}}]$\\
   \hline
   \begin{tabular}[c]{@{}l@{}}N\\ NH \\NH$\Delta$ \end{tabular}&1& $0.4$ &\begin{tabular}[c]{@{}l@{}}$2.44$\\ $2.33$\\2.33 \end{tabular}& \begin{tabular}[c]{@{}l@{}}$2.81$\\ $2.65$\\2.65 \end{tabular}& \begin{tabular}[c]{@{}l@{}}$12.33$\\ $12.38$\\12.39 \end{tabular}& \begin{tabular}[c]{@{}l@{}}$1070$\\ $1078$\\1072 \end{tabular}& \begin{tabular}[c]{@{}l@{}}5.26\\ 5.36\\ 5.36\end{tabular}& \begin{tabular}[c]{@{}l@{}}30.08\\ 21.16\\ 21.13\end{tabular}&\begin{tabular}[c]{@{}l@{}}---\\ ---\\ ---\end{tabular}& \begin{tabular}[c]{@{}l@{}}---\\ ---\\ ---\end{tabular}&\begin{tabular}[c]{@{}l@{}}---\\ ---\\ ---\end{tabular}\\
   \hline
    \begin{tabular}[c]{@{}l@{}}N\\ NH \\NH$\Delta$ \end{tabular}&2 & $0.2$ & \begin{tabular}[c]{@{}l@{}}$2.49$\\ $2.28$ \\ 2.29 \end{tabular} & \begin{tabular}[c]{@{}l@{}}$2.89$\\ $2.61$\\2.62 \end{tabular}& \begin{tabular}[c]{@{}l@{}}$12.83$\\ $12.56$\\ 12.52 \end{tabular}& \begin{tabular}[c]{@{}l@{}}$1007$\\ $1070$ \\ 1073 \end{tabular}& \begin{tabular}[c]{@{}l@{}}5.00\\ 5.42\\ 5.42 \end{tabular}& \begin{tabular}[c]{@{}l@{}}71.83\\ 46.42 \\ 45.41\end{tabular}&\begin{tabular}[c]{@{}l@{}}---\\ ---\\ ---\end{tabular}&\begin{tabular}[c]{@{}l@{}}---\\ ---\\ ---\end{tabular}&\begin{tabular}[c]{@{}l@{}}---\\ ---\\ ---\end{tabular}\\
   \hline
    \begin{tabular}[c]{@{}l@{}}N\\ NH \\NH$\Delta$ \end{tabular}&2 &$Y_{\nu e}= 0$ &\begin{tabular}[c]{@{}l@{}}$2.43$\\ $2.22$\\ $2.22$ \end{tabular} & \begin{tabular}[c]{@{}l@{}}$2.83$\\ $2.54$\\ $2.55$ \end{tabular}& \begin{tabular}[c]{@{}l@{}}$13.66$\\ $13.03$\\ $12.86$\end{tabular}& \begin{tabular}[c]{@{}l@{}}$1068$\\ $1084$\\ $1107$\end{tabular}& \begin{tabular}[c]{@{}l@{}}5.59\\ 5.56\\ 5.65 \end{tabular}& \begin{tabular}[c]{@{}l@{}}23.08\\51.22\\49.41 \end{tabular}&\begin{tabular}[c]{@{}l@{}}62.61\\ 48.90\\47.30 \end{tabular}& \begin{tabular}[c]{@{}l@{}}1470\\ 1510\\ 1520\end{tabular}&\begin{tabular}[c]{@{}l@{}}238\\ 280\\ 290\end{tabular}\\ 
   \hline
    \begin{tabular}[c]{@{}l@{}}N\\ NH \\NH$\Delta$ \end{tabular}&$T = 0$ & $Y_{\nu e}= 0$ & \begin{tabular}[c]{@{}l@{}}$2.26$\\ $2.21$\\$2.21$ \end{tabular}& \begin{tabular}[c]{@{}l@{}}$2.67$\\ $2.60$\\ $2.61$\end{tabular}& \begin{tabular}[c]{@{}l@{}}$13.04$\\ $12.65$\\$12.55$\end{tabular}& \begin{tabular}[c]{@{}l@{}}$611$\\ $876$\\$1021$ \end{tabular}& \begin{tabular}[c]{@{}l@{}}3.56 \\ 4.87\\5.46 \end{tabular}& \begin{tabular}[c]{@{}l@{}}{---}\\ {---}\\--- \end{tabular}&\begin{tabular}[c]{@{}l@{}}---\\ ---\\ ---\end{tabular}& \begin{tabular}[c]{@{}l@{}}1425\\ 1476\\ 1486\end{tabular}&\begin{tabular}[c]{@{}l@{}}177\\ 222\\ 230\end{tabular}\\
  \hline
  \multicolumn{12}{|c|}{Fixed $M_b$ and the corresponding stellar matter properties} \\
  \hline
    Particle content &S$/$n$_B$ &Y$_{L,e}$ & $M_b[{\rm M}_{\odot}]$& M$[{\rm M}_{\odot}]$ & $R[{\rm km}]$&$\varepsilon_0[{\rm MeV fm^{-3}}]$&$n_c[n_0]$& $T_c[{\rm MeV}]$ & $T_C[{\rm MeV}]$& $\mu_c[{\rm MeV}]$&$P_c[{\rm MeV fm^{-3}}]$\\
   \hline
   \begin{tabular}[c]{@{}l@{}}N\\ NH \\NH$\Delta$ \end{tabular}&1& $0.4$ &\begin{tabular}[c]{@{}l@{}}$2.45(1.57)$ \end{tabular}& \begin{tabular}[c]{@{}l@{}}$2.19(1.48)$\\ $2.18(1.47)$\\2.19(1.47) \end{tabular}& \begin{tabular}[c]{@{}l@{}}$13.65(14.55)$\\ $13.49(14.55)$\\13.46(14.545) \end{tabular}& \begin{tabular}[c]{@{}l@{}}$569.34(348.42)$\\ $631.04(348.38)$\\638.30(348.38) \end{tabular}& \begin{tabular}[c]{@{}l@{}}3.30(2.19)\\3.59(2.19)\\3.62(2.19) \end{tabular}& \begin{tabular}[c]{@{}l@{}}24.21(18.71)\\ 20.47(18.26)\\20.24(18.15) \end{tabular}&\begin{tabular}[c]{@{}l@{}}---\\ ---\\ ---\end{tabular}&\begin{tabular}[c]{@{}l@{}}---\\ ---\\ ---\end{tabular}&\begin{tabular}[c]{@{}l@{}}---\\ ---\\ ---\end{tabular}\\
   \hline
   \begin{tabular}[c]{@{}l@{}}N\\ NH \\NH$\Delta$ \end{tabular}&2 & $0.2$ & \begin{tabular}[c]{@{}l@{}}$2.45(1.57)$ \end{tabular} & \begin{tabular}[c]{@{}l@{}}$2.19(1.46)$\\ $2.17(1.48)$\\ 2.18(1.49) \end{tabular}& \begin{tabular}[c]{@{}l@{}}$14.44(15.57)$\\ $13.71(15.25)$\\13.60(15.12)\end{tabular}& \begin{tabular}[c]{@{}l@{}}$501.81(301.75)$\\ $653.83(329.52)$\\659.96(340.48) \end{tabular}& \begin{tabular}[c]{@{}l@{}}3.01(1.93)\\ 3.72(2.09)\\3.76(2.16) \end{tabular}& \begin{tabular}[c]{@{}l@{}}57.00(41.93)\\ 43.99(38.36)\\42.66(36.96) \end{tabular}&\begin{tabular}[c]{@{}l@{}}---\\ ---\\ ---\end{tabular}&\begin{tabular}[c]{@{}l@{}}---\\ ---\\ ---\end{tabular}&\begin{tabular}[c]{@{}l@{}}---\\ ---\\ ---\end{tabular}\\
   \hline
   \begin{tabular}[c]{@{}l@{}}N\\ NH \\NH$\Delta$\end{tabular}&2 &$Y_{\nu e}= 0$ &\begin{tabular}[c]{@{}l@{}}$2.45(1.57)$ \end{tabular} & \begin{tabular}[c]{@{}l@{}}$2.17(1.48)$\\ $2.16(1.46)$\\ 2.16(1.47) \end{tabular}& \begin{tabular}[c]{@{}l@{}}$14.50(15.56)$\\ $13.50(15.41)$\\13.36(15.027)\end{tabular}& \begin{tabular}[c]{@{}l@{}}$491.47(305.51)$\\ $720(333.67)$\\710(354.47)\end{tabular}& \begin{tabular}[c]{@{}l@{}}2.94(1.96)\\4.08(2.06)\\4.02(2.26) \end{tabular}& \begin{tabular}[c]{@{}l@{}}54.89(43.27)\\46.74(38.59)\\45.68(38.25) \end{tabular}&\begin{tabular}[c]{@{}l@{}}---\\ ---\\ ---\end{tabular}&\begin{tabular}[c]{@{}l@{}}---\\ ---\\ ---\end{tabular}&\begin{tabular}[c]{@{}l@{}}---\\ ---\\ ---\end{tabular}\\ 
   \hline
   \begin{tabular}[c]{@{}l@{}}N\\ NH \\NH$\Delta$ \end{tabular}&$T = 0$ & $Y_{\nu e}= 0$ & \begin{tabular}[c]{@{}l@{}}$2.45(1.57)$ \end{tabular}& \begin{tabular}[c]{@{}l@{}}$2.10(1.42)$\\2.10(1.41)\\2.10(1.43)\end{tabular}& \begin{tabular}[c]{@{}l@{}}$13.22(13.22)$\\13.00(13.21)\\12.85(13.14)\end{tabular}& \begin{tabular}[c]{@{}l@{}}$503.68(343.21)$ \\607.57(341.77)\\611.71(363)\end{tabular}& \begin{tabular}[c]{@{}l@{}}3.14(2.29)\\ 3.66(2.26)\\3.69(2.39)\end{tabular}& \begin{tabular}[c]{@{}l@{}}{---}\\ {---} \\---\end{tabular}&\begin{tabular}[c]{@{}l@{}}---\\ ---\\ ---\end{tabular}&\begin{tabular}[c]{@{}l@{}}---\\ ---\\ ---\end{tabular}&\begin{tabular}[c]{@{}l@{}}---\\ ---\\ ---\end{tabular}\\
  \hline
\end{tabular}
\caption{ 
{Here, N represents nucleons only, NH represents nucleon plus hyperons admixture, and NH$\Delta$ represents nucleon plus hyperons plus $\Delta$-isobars admixture.} We show the maximum gravitational mass (M$_{\rm max}$) and its corresponding radii ($R$), the central energy density ($\varepsilon_0$), the central baryon density $n_c$, the core temperature ($T_{co}$) and the critical temperature ($T_C$) for the evolution stages of the PNS considered. The stars without $P_c$ and $\mu_c$ values do not undergo a phase transition to a free quark core. Additionally, we choose two fixed baryonic masses; ${\rm M_b}=1.57\,{\rm M}_{\odot}$ and ${\rm M_b}=2.45\,{\rm M}_{\odot}$ and calculate the other properties of the stellar matter for the four stages of the stellar evolution. The results are arranged such that the star with the smaller baryonic mass is in a bracket and the larger one is out of the bracket i.e. 2.45(1.57). 
} 
\label{T1a}
\end{center}
\end{table*}

On Tab.~\ref{T1a}, we present the results of the stellar properties determined from our calculations. Upon examination of the table, we observe that the maximum stellar masses generally decrease as the degrees of freedom of the stellar matter increase, as expected. Additionally, from the neutrino-trapped region where $Y_{L,e}$ is fixed, we note that the maximum stellar mass decreases slightly as the star deleptonizes, except for stars composed solely of nucleons, which exhibit no significant changes in mass which may be due to accretion. However, during deleptonization, the stellar radii increase alongside higher core temperatures as neutrinos diffuse from the stellar core. For the neutrino transparent matter, on the other hand, the maximum stellar mass and radii decrease as the star cools down and shrinks through thermal emissions. For further clarity, we profiled two stars with fixed baryon masses $\rm M_b = 1.57\,\rm M_\odot$ and $\rm M_b = 2.45\,\rm M_\odot$ to determine their corresponding gravitational masses, radii, core temperature, central energy, and baryon densities and presented them on the table. 

We settled on the fixed $\rm M_b$ because it is one of the quantities that is expected to remain conserved during stellar evolution, unlike the gravitational mass and radii that depend largely on the lepton fraction and consequently the neutrino diffusion and thermal radiations. Although the emergence of quark cores impacts $\rm M_b$, this effect is minimal since the transformation also conserves baryon mass; therefore, the fixed baryon mass remains a reasonable choice for ensuring that we are profiling the same star at all stages. We also ensure that our choice satisfies the 2M$\odot$ threshold and the 1.4M$\odot$; measured for pulsars based on data from the NICER observatory.

Furthermore, the baryon mass is expected to remain approximately conserved in an isolated NS; however, this can change depending on the evolutionary history of the star. In particular, if the star accretes mass (which could lead to black hole formation) or if an internal phase transition occurs, resulting in the formation of exotic baryons and/or deconfined quark cores, then $\rm M_b$ can increase, as shown in Tab.~\ref{T1a}. This process affects the internal structural formations and the gravitational mass of the star; however, the total baryon count must remain conserved in all cases \cite{Prakash:1996xs, Pons:1998mm}.

\begin{table}[ht]
\begin{center}
\begin{tabular}{|c |c| c| c| c| c| }
 \hline
 \multicolumn{6}{|c|}{Quark cores} \\
  \hline
  Particles & S$/$n$_B$, Y$_{\nu_e}$ & $M_H[{\rm M}_{\odot}]$& $M_Q[{\rm M}_{\odot}]$ & $Q_c$[\%]& $Q_{\rm stb}$[\%] \\
   \hline
   \begin{tabular}[c]{@{}l@{}}N\\ NH\\ NH$\Delta$ \end{tabular}& $2;\;0$ &\begin{tabular}[c]{@{}l@{}}2.40\\ 2.22\\2.22\end{tabular}& \begin{tabular}[c]{@{}l@{}}0.54\\ 0.46\\0.34\end{tabular}& \begin{tabular}[c]{@{}l@{}}18\\ 17\\13 \end{tabular}& \begin{tabular}[c]{@{}l@{}}1.2\\ 0\\0\end{tabular}\\
   \hline
   \begin{tabular}[c]{@{}l@{}}N\\ NH\\ NH$\Delta$ \end{tabular} & $T=0;\; 0$ & \begin{tabular}[c]{@{}l@{}}2.23\\2.20\\2.20 \end{tabular}& \begin{tabular}[c]{@{}l@{}} 0.57\\ 0.69\\0.54\end{tabular}& \begin{tabular}[c]{@{}l@{}}24\\21\\20\end{tabular}& \begin{tabular}[c]{@{}l@{}}1.3\\ 0.45\\0.45 \end{tabular}\\
  \hline
\end{tabular}
\caption{The proportion of QM configuration in the stellar matter. From the table above; $\rm M_H$ represents the maximum stellar mass in the hadronic matter phase, $\rm M_Q$ is the maximum stellar mass in the QM phase, $Q_c$ is the QM core in the entire hybrid star configuration, and $Q_{\rm stb}$ is the quark core in the stable hybrid star configuration up to a stable star with maximum mass $\rm M_{max}$.} 
\label{Tt}
\end{center}
\end{table}
In Table~\ref{Tt}, we determined \( Q_c \) based on the values of \( \rm M_{\rm H} \) and \( \rm M_{\rm Q} \) under the same thermodynamic conditions. To calculate \( Q_{\rm stb} \), we assumed that all hadronic matter contributes to forming a stable stellar configuration with a maximum mass, \( \rm M_{max} \). Therefore, the difference between \( \rm M_{max} \) and \( \rm M_H \) represents the quark core present in this stable stellar configuration. We adopt this approach for two reasons: First, none of the values of $\rm M_H$ in the table above exceed the corresponding $\rm M_{\rm max}$ values in Tab.~\ref{T1a}. Second, hadronic matter constitutes the low-density region of the hybrid EoS. From the table above, we can infer that the QM core decreases in size when heavier baryons are introduced into stellar matter. Additionally, there are quark cores in the stable stellar configuration at $T=0$ for all particle combinations (N, NH, and NH$\Delta$) in different proportions in contrast to the hot we find a hybrid star with nucleonic matter composition only.

\begin{figure*}[ht!]
  \includegraphics[scale=0.5]{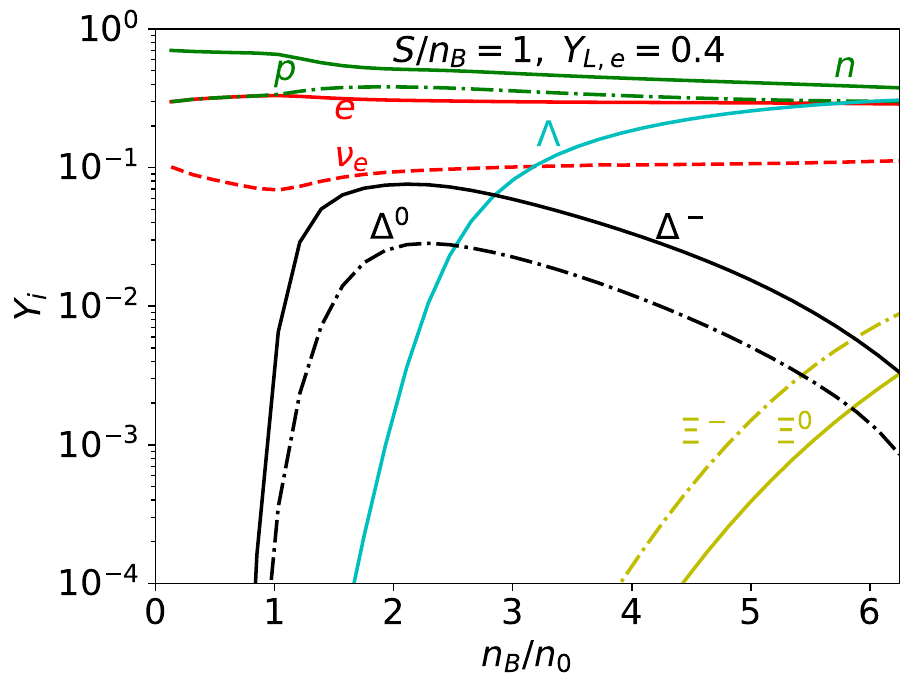}
  \quad
   \includegraphics[scale=0.5]{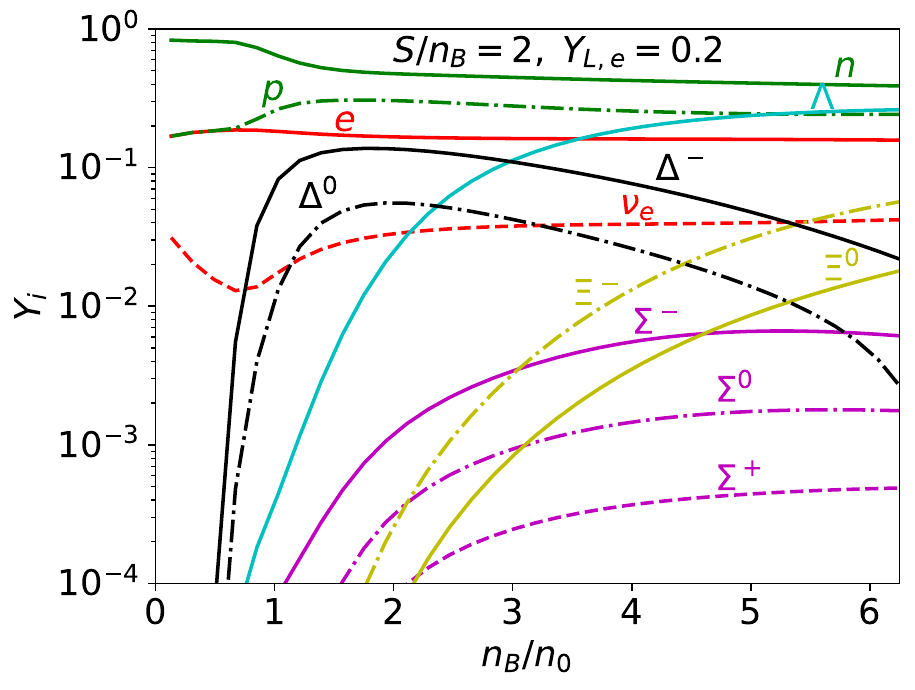}
  \quad
  \includegraphics[scale=0.5]{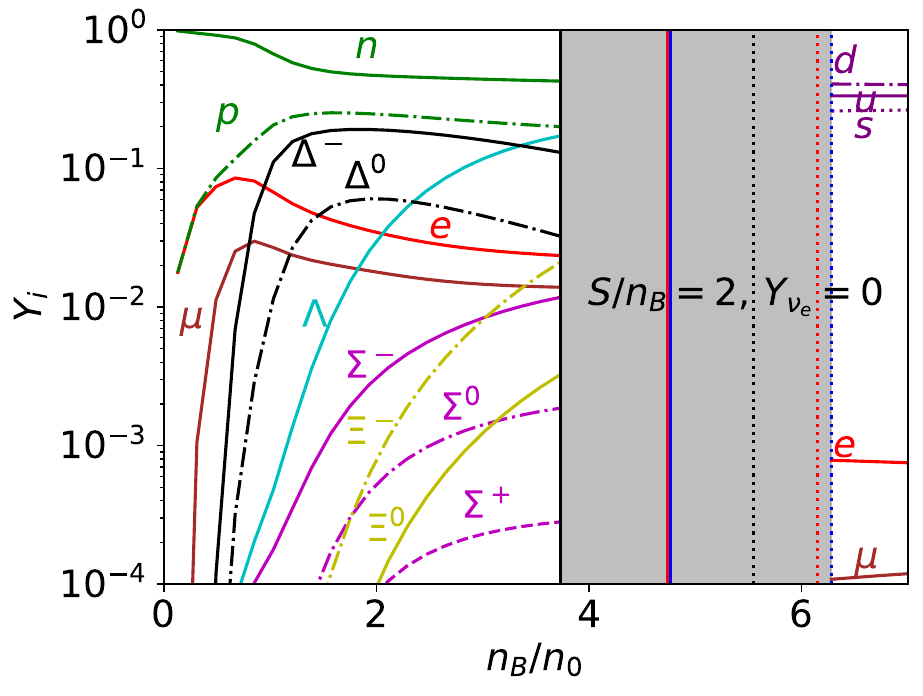}
  \quad
 \includegraphics[scale=0.5]{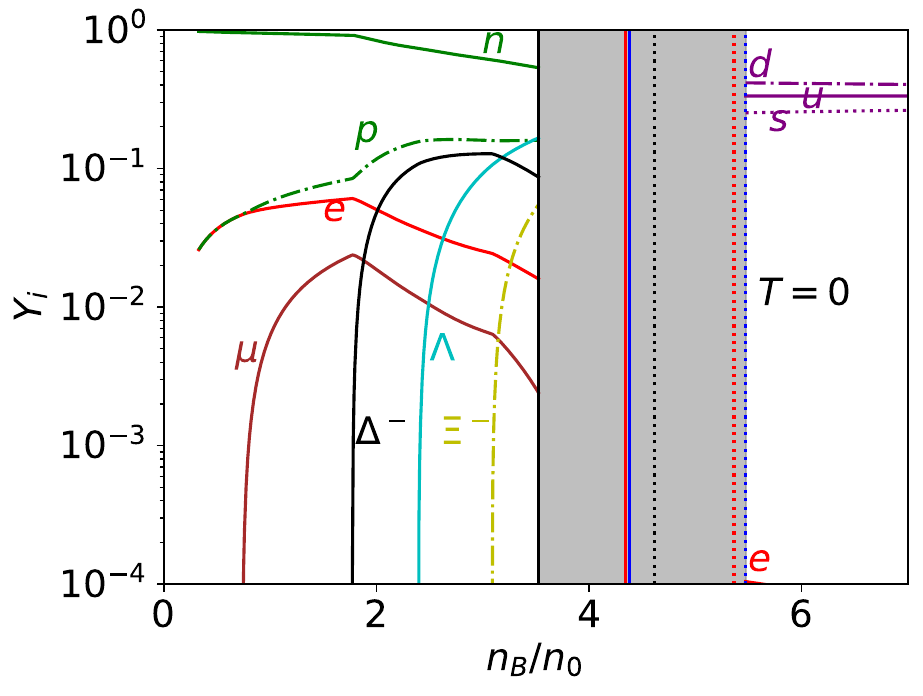}
\caption{The particle fraction for the PNS from birth to maturity as a cold catalyzed hybrid NS. The upper panels show the stages when the star is neutrino-rich and the lower panels when it is neutrino-poor. The proto-neutron stars undergo a phase transition from a hadronic to a quark matter phase once all the neutrinos have escaped from the stellar core. They then continue cooling down to form a cold hybrid NS, as depicted in the bottom panels. The solid vertical lines in the lower panels show when the phase transition of each particle begins and the corresponding dot-dot line shows where the phase transition ends for each particle. The black lines represent the nucleon, the red lines represent nucleon plus hyperons and the blue line represents the nucleon plus hyperons plus $\Delta$-isobars.}
    \label{pfs}
\end{figure*}

From the snapshots of the particle distributions depicted in Fig.~\ref{pfs}, two fates await the supernova remnants at each stage of its evolution: it can either form a black hole or a PNS. However, the formation of a black hole signifies the culmination of stellar evolution. Therefore, in this study, we presume that all conditions favored the formation of PNS \cite{Prakash:1996xs}. In the first stage ($S/n_B=1$ and $Y_{L,e}=0.4$), if the explosion lacks the requisite strength to deleptonize the mantle of the star, it promptly accretes matter, leading to the formation of a black hole. In such a scenario, the evolution of the star is abruptly terminated, and it does not progress to the second stage ($S/n_B=2$ and $Y_{L,e}=0.2$). During the transition from the second to the third stage ($S/n_B=2$ and $Y_{\nu_e}=0$), the star undergoes deleptonization via neutrino diffusion. If neutrino loss transpires too rapidly, it can induce softening of the EoS, thereby increasing the risk of black hole formation, particularly if the gravitational pressure reaches sufficiently high levels. In the third stage, once all neutrinos have escaped from the stellar matter, the likelihood of strangeness-carrying hyperons emerging in the stellar matter rises. If the formation of strangeness-carrying hyperonic matter becomes excessive, it can potentially lead to the formation of a black hole \cite{Vidana:2002rg, Burgio2007}. This phenomenon occurs if there is an accretion of matter at this stage.

In Fig.~\ref{pfs}, we illustrate the particle distribution in the stellar matter determined using (\ref{yi}). The particle distribution provides valuable insights into the isospin asymmetry of the stellar matter as the star evolves \cite{Issifu:2023qyi, Chu:2017huf}, highlighting how temperature influences the emergence of particles during stellar evolution and its consequent impact on the EoS \cite{Raduta:2020fdn}. The isospin asymmetry of the quark matter is computed using
\begin{equation}
\delta_q = 3\dfrac{n_d-n_u}{n_d+n_u},
\end{equation}
where $n_3 = n_d-n_u$ represents the isospin density and $n_B=(n_d+n_u)/3$ denotes the baryon density for a two-flavor quark system. Conversely, the isospin asymmetry of the hadronic matter is determined by
\begin{equation}
\delta_H = \dfrac{n_n-n_p}{n_n+n_p},
\end{equation}
where $n_B = n_n+n_p$. Along the panels, it is clear that the ratio of the particle fraction $Y_p/Y_n$ decreases along the panel as the star evolves. This means the isospin asymmetry increases along the evolution lines. However, the variations in $Y_u/Y_d$ are not evident from the bottom panels, as the quark fractions appear to remain constant at higher densities, well beyond several times $n_0$. Nonetheless, as discussed in \cite{Issifu:2023qoo}, it was noted that this ratio decreases as PSSs evolve; consequently, the isospin asymmetry $\delta_q$ increases with PSS evolution. Additionally, there is a delay in the emergence of strangeness-carrying particles, primarily the $\Lambda$ hyperon, in the low entropy ($S/n_B=1$ and $Y_{L,e}=0.4$, corresponding to low temperature) neutron-rich matter and in neutrino-poor stellar matter with $T=0$, as illustrated in the first and fourth panels, respectively. As the matter deleptonizes and $S/n_B$ increases ($S/n_B=2$ and $Y_{L,e}=0.2$), the strangeness-carrying particles shift further towards the low-density region. Subsequently, after all the neutrinos have escaped from the core ($S/n_B=2$ and $Y_{\nu_e}=0$), all the strangeness-carrying particles appear either before or at $n_B\approx 2n_0$, as evident in the second and third panels. {This establishes a relation between the appearance of strangeness-carrying particles with the increase in temperature and entropy.} Another crucial point to note is that the presence of neutrinos hinders the appearance of free quarks in stellar matter.

Moreover, the $\Delta^-$ heavy non-strange baryon prominently appears in the stellar matter at each stage along the evolution lines, causing a reduction in the lepton population. The most prevalent strangeness-carrying hyperon in the stellar matter across all regimes is the $\Lambda$. At low $S/n_B$ (first panel) and low temperatures (fourth panel), $\Lambda$ and $\Xi$ hyperons emerge as the dominant strangeness-carrying particles at low $n_B$. Nonetheless, at higher $S/n_B$, this phenomenon changes, and $\Lambda$ and $\Sigma$ strangeness-carrying hyperons dominate at low $n_B$, as observed in the second and third panels. {Hence, the temperature favors the production of $\Sigma$ particles, as evident in the second and third panels. This is consistent with the finding in \cite{Sedrakian:2022kgj, Raduta:2020fdn} where the authors study PNSs with exotic baryon content and hot quark matter and proto-neutron stars in \cite{PhysRevC.100.015803} with exotic baryons as well. Therefore, the $\Sigma$ particles do not appear in $T=0$ and $S/n_B =1,\, Y_{L,e}=0.4$; whose core temperature is small.} Zooming in on the bottom panels, we observe a phase transition from hadronic to quark matter phase. The phase transition begins lowly immediately after the neutrinos have completely escaped from the stellar core, at the point where the matter is expected to be maximally heated. This is evident by the wider gray area in the third panel compared to the fourth one. In the fourth panel, it's evident that the quark core expands significantly as the matter cools down to $T=0$, with comparatively faster phase transitions. We have depicted the onset of phase transitions with solid vertical lines and the cessation points with a corresponding dotted vertical line. As the degrees of freedom of the stellar matter increase, the phase transition slows down.

\begin{figure*}[ht!]
  \includegraphics[scale=0.5]{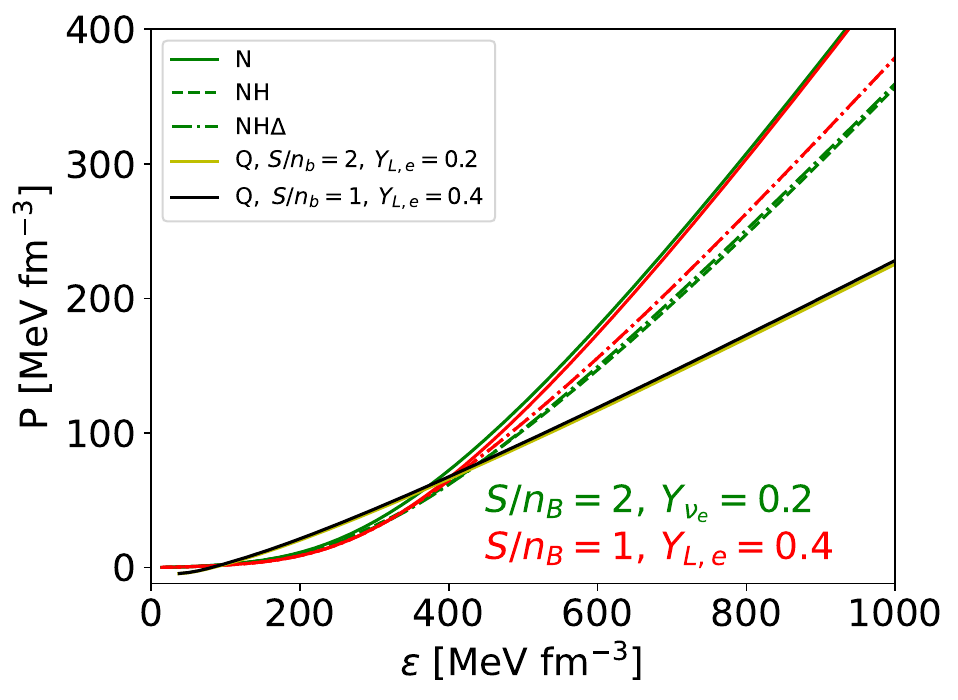}
  \quad
   \includegraphics[scale=0.5]{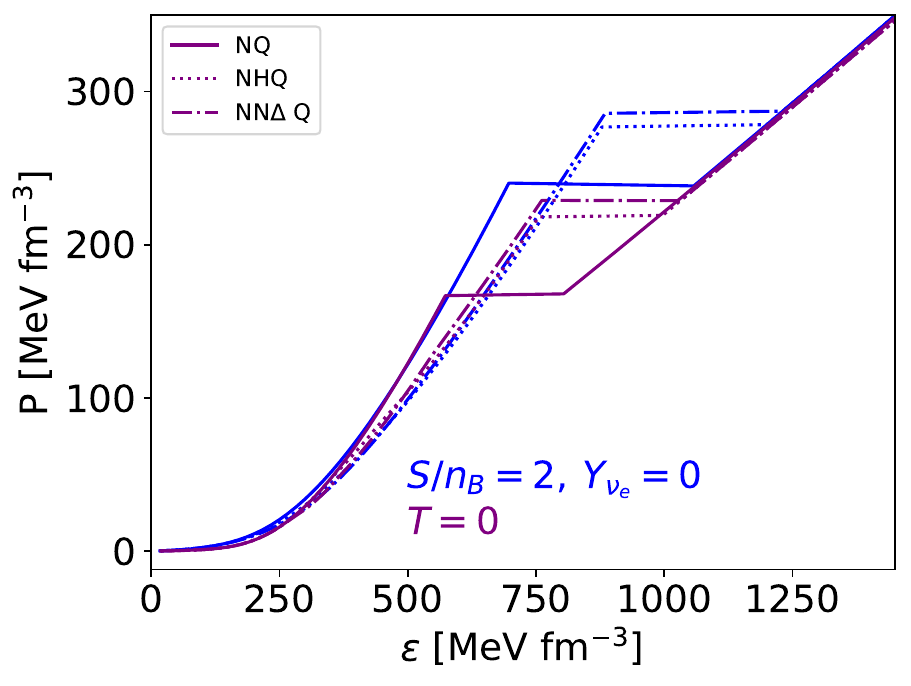}
  \caption{The figures display the pressure as a function of the energy density. The left panel shows the EoSs of the stellar matter when the matter is neutrino-rich with no phase transitions. {The right panel shows both a neutrino-free hybrid PNS and a cold hybrid star EoS, each exhibiting a phase transition.}}
    \label{eos}
\end{figure*}

In Fig.~\ref{eos}, we show the EoSs for various stages of the star's evolution. In the left panel, we show the EoSs of both the hadronic and quark phases when neutrinos are trapped in the stellar core. We determined from Fig.~\ref{equ} that there is no chemical and mechanical equilibrium established between both matter phases at this regime hence, there is no hadron-quark phase transition here. 
On the other hand, the curves in the right panel show the EoSs for neutrino-transparent matter where phase transition from hadronic to quark matter phase occurs. This is evident by the discontinuity in the energy density. Generally, we observe that the EoSs become softer as the degrees of freedom of the stellar matter increase as expected \cite{Ribes:2019kno, PhysRevC.98.045801}. This property is essential in the determination of the structure of the corresponding stars since stiffer EoS signifies a higher maximum gravitational mass of the star and vice versa. In the right panel, it is noticeable that the phase transition from hadronic to quark matter phase for the fixed entropy stars ($S/n_B=2,\; Y_{\nu_e}=0$) happens at higher energy densities compared to the $T=0$ stars. This is attributed to the slow nature of the phase transition in the fixed entropy stars compared to the cold ones as discussed in Fig.~\ref{pfs}. 

In the neutrino-trapped regime, higher fixed $Y_{L,e}$ results in the 
{increase of the} stiffness of the EoS, hence the first stage of the star's evolution is associated with stiffer EoSs compared to the deleptonization phase in the second stage where the entropy increases and $Y_{L,e}$ decreases. This correlation can be attributed to the emergence of strangeness-carrying hyperons in the stellar matter, as depicted in Fig.~\ref{pfs}. The increased presence of hyperons at lower densities results in a softer EoS in this regime. In the neutrino-transparent matter, where the hybrid stars are formed, the EoSs become softer as the star cools down from $S/n_B=2$, undergoing catalyzation at $S/n_B=0$, and reaches $T=0$ when a 'mature' NS is formed.

\begin{figure}[ht!]
  \includegraphics[scale=0.45]{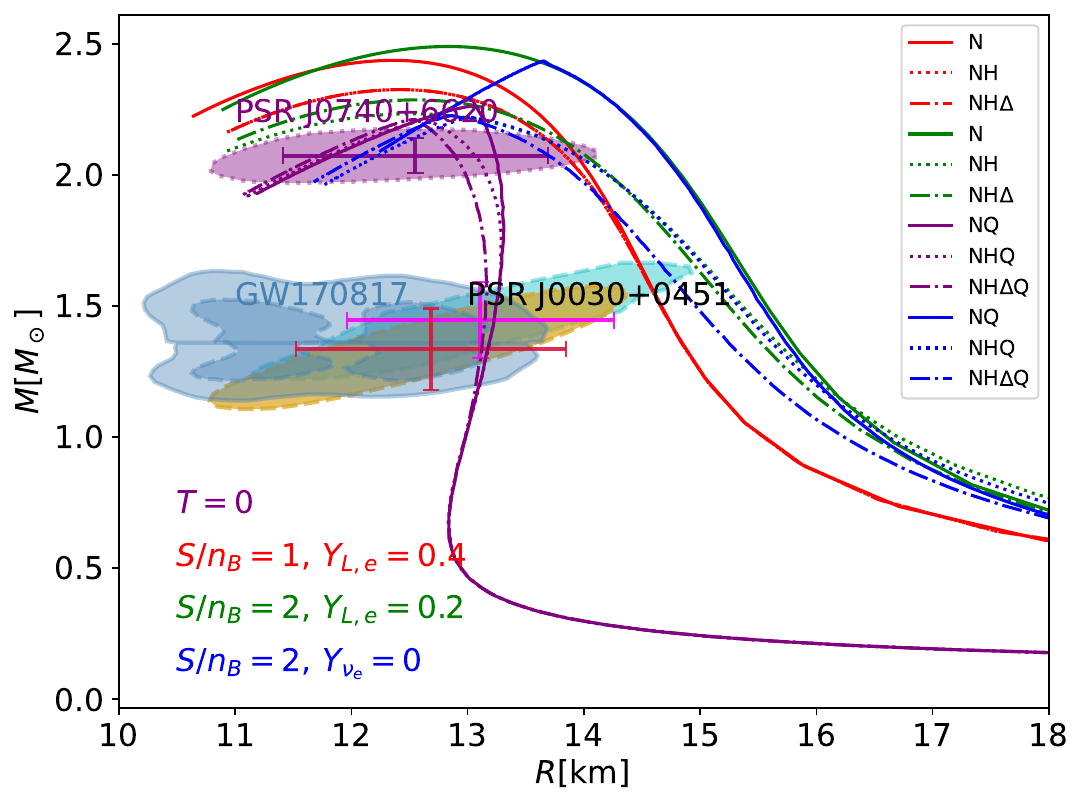}
\caption{Here we show the gravitational masses of the stars as a function of radii. The contours marked on the graph are the observed pulsars determined at various confidence levels 
to show that the EoS at $T=0$ satisfies these observational data. The purple contour represents the PSR J0740+6620 pulsar with its radius margin at a 68\% confidence level (CL) \cite{Riley:2021pdl, Miller:2021qha}. The PSR J0030+0451 pulsar is indicated in cyan \cite{Miller:2019cac} and goldenrod \cite{Riley:2019yda} for the 68\%CL measurement by two groups, with their respective radii and error margins shown in magenta and crimson. The GW170817 region is highlighted in steel blue, with the outer contour representing the 90\% CL \cite{LIGOScientific:2018cki, LIGOScientific:2017vwq} and the inner contour indicating the 50\% CL \cite{LIGOScientific:2017vwq} regions. }
    \label{mr}
\end{figure}
The mass-radius diagram that represents the macroscopic structure of the NSs is determined by solving the Tolman–Oppenheimer–Volkoff (TOV) equations \cite{PhysRev.55.374}, assuming a hydrostatic spherically symmetric fluid. The relevant equations to solve are 
\begin{align}
    \frac{dP(r)}{dr}&=-[\varepsilon(r) + P(r)]\frac{M(r)+4\pi r^3 P(r)}{r^2-2M(r)r}, \label{eq1}\\
    \frac{dM(r)}{dr}&=4\pi r^2 \varepsilon(r).\label{eq2}
\end{align}
With the radial coordinate $r$, gravitational mass as a function of radius $M(r)$, pressure $P(r)$, and energy density $\varepsilon(r)$, we utilized natural units where $G=c=1$. Fig.~\ref{mr} depicts the solution of these equations. The baryon mass ($M_B$) is computed using
\begin{equation}
    \dfrac{dM_B(r)}{dr} =4\pi  m_n \dfrac{r^2 n_B(r)}{[1-2M(r)/r]^{1/2}},
\end{equation}
where $m_n=939\,\rm MeV$ is the nucleon mass.

At this juncture, it's crucial to note that there are two broader approaches to investigating PNSs: the quasi-static approximation description, applicable when the stellar matter is isentropic, and the full hydrostatic simulation, utilized when the stellar matter is out of equilibrium. Thus, the TOV equations are applicable in the former, not the latter (see e.g. \cite{Raduta:2020fdn, Oertel:2016xsn, Sedrakian:2022kgj, Malfatti:2019tpg, Shao:2011nu, Chen:2021edy, Kumari:2021tik}). In the quasi-static approximation description the time derivatives of density, pressure, and the metric vanish therefore, the average hydrostatics effects that focus on the evolution of intensive thermodynamic quantities such as the $Y_{L,e}$ and the $S/n_B$ over the Kelvin-Helmholtz timescale is enough for the analysis \cite{Pons:1998mm, roberts2012new}. The boundary conditions at this stage are $P(R)=P_{\rm surf}$, with $P_{\rm surf}$, the pressure on the surface of the star, and $R$ is the stellar radius, in the case of `mature' NSs at $T=0$, $P(R)=0$. In the finite temperature (fixed entropy) scenario, the selection of $P_{\rm surf}$ is somewhat arbitrary owing to the presence of thermal pressure. Nevertheless, its value has been found to impact the properties of the stellar mantle during its initial formation stages. However, studies have shown that significantly small values of $P_{\rm surf}$ have a negligible effect on the long-term evolution of the star's internal structures \cite{RevModPhys.74.1015, Gulminelli:2015csa}.

The outcomes of the mass-radius diagram derived for pure nucleonic matter, nucleons plus hyperons, and nucleons plus hyperons plus $\Delta$-isobars admixed hypernuclear matter all adhere to the $2\,\rm M_\odot$ mass constraint imposed on NSs. As additional degrees of freedom are introduced into the stellar matter, the maximum masses decrease, in accordance with expectations. The observed radii also fall within the ranges determined for the recently observed pulsar from the NICER observatory: PSR J0740+6620 \cite{NANOGrav:2019jur, Fonseca:2021wxt}.

\begin{figure}[ht!]
  \includegraphics[scale=0.5]{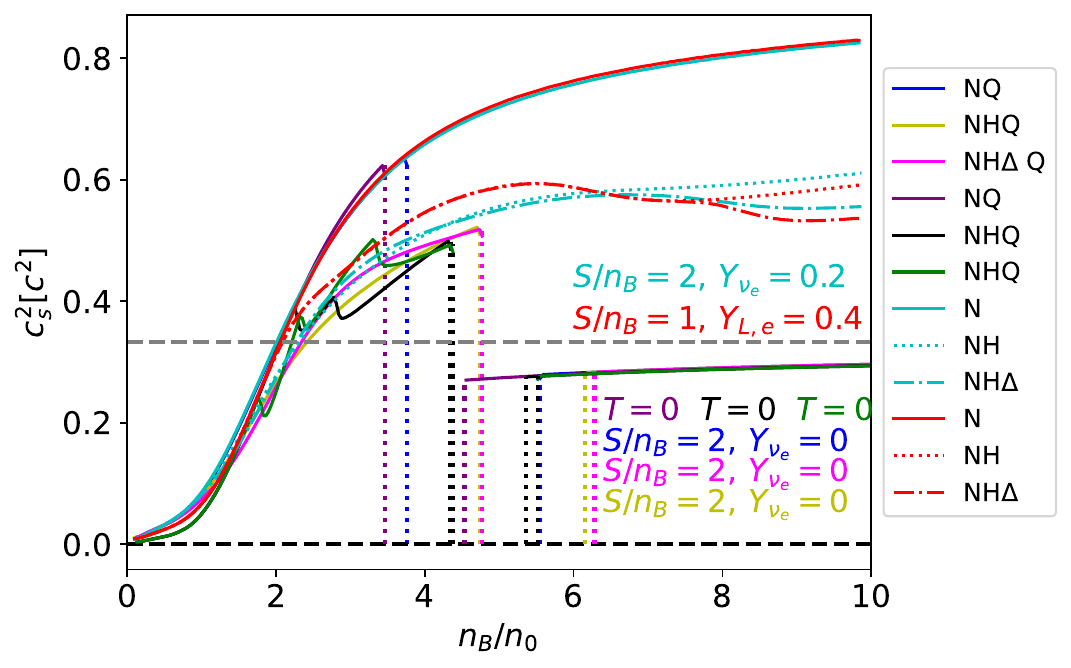}
\caption{The figure shows the velocity of sound squared as a function of the baryon density. The horizontal gray line is the conformal limit, $c_s^2=1/3$, below this line the stars are made up of self-bound quarks, and above it, we have hadronic stars. The black horizontal line is the $c_s^2=0$ line. The distance between two vertical dotted lines of the same kind, which touches the $c_s^2=0$ line, indicates the region where the phase transition occurs. The curves increase monotonically with the baryon density, however, at the point of phase transition the $c_s$ drops drastically below the conformal limit.}
    \label{sound}
\end{figure}

 The square of the sound velocity depicted in Fig.~\ref{sound} is computed using the relation
\begin{equation}
c_s^2 = \dfrac{\partial P}{\partial \varepsilon},
\end{equation}
which assists in classifying the particle composition of NSs using conformal symmetry arguments and the EoS as input. The $c_s^2$ leverages the distinct properties exhibited by QCD theory in both perturbative and nonperturbative regimes to differentiate between particle compositions. Studies indicate that in the perturbative regime of QCD theory particles exhibit behavior akin to scale invariance, whereas the opposite behavior is observed in the nonperturbative regime. This is distinguishable using $c_s^2$ since the matter is exactly scale-invariant at $c_s^2 = 1/3$. This value is approached from below at higher density regions ($n_B > 40n_0$) in QM \cite{Kurkela:2009gj}, where matter is considered to be scale-invariant. However, in the hadronic matter, $c_s^2$ rises to values $c_s^2 \gtrsim 0.5$, clearly violating the scale-invariant limit. The $c_s^2$ must also satisfy causality, $c_s^2\,\leq\,1$, and thermodynamic stability,  $c_s^2>0$, conditions simultaneously \cite{Annala:2019puf, Bedaque:2014sqa}. We observe from Fig.~\ref{sound} that the curves for the neutrino-trapped matter grow monotonically beyond the conformal limit without decreasing. On the other hand, in neutrino-transparent matter, where hadron-quark phase transitions take place, the $c_s^2$ curves rise with density in the hadronic regime until they reach their maximum. At densities where free quarks begin to appear in the stellar matter, $c_s^2$ drops quickly below the conformal limit, indicating the presence of conformal invariant matter \cite{Fraga:2013qra, Borsanyi:2010cj}. The region of the steep drop in the curves represents the phase transition region, similar to the regions marked with gray areas in Fig.~\ref{pfs}. The bumps in the $T=0$ curves (purple)  may represent shocks propagating through stacks of different particles. { Comparing with the $n_B/n_0$ at which new particles appear in Fig.~\ref{pfs}. The $\Delta^-$ appears at $\sim 1.8n_0$, $\Lambda^0$ at $\sim 2.2n_0$ and the $\Xi^-$ at $\sim 3.3n_0$ in the same position the bumps appear depending on the matter composition. For NH matter, we have two bumps: the appearance of $\Lambda^0$ and $\Xi^-$ hyperons at $\sim 2.2n_0$ and $\sim 3.3n_0$ respectively. In the NH$\Delta$ matter we have three bumps first at $\sim 1.8n_0$ when $\Delta^-$ appears, the second at $\sim 2.2n_0$ and the last at $\sim 3.3n_0$ when $\Xi^-$ appears. Thus, no bumps are seen when matter is composed of only nucleons.} 

\begin{figure}[ht!]
  \includegraphics[scale=0.5]{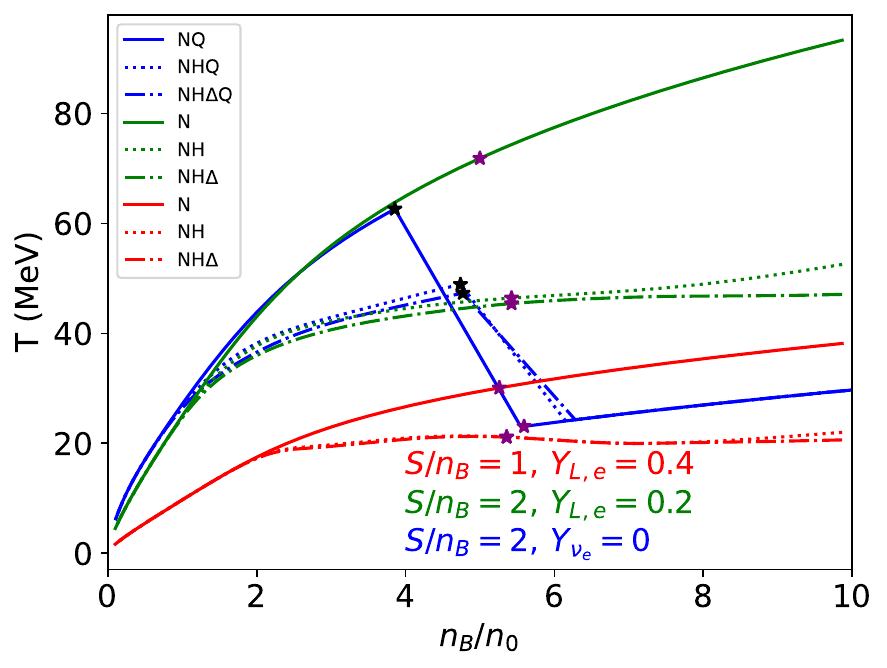}
\caption{The graph illustrates the temperature of the stellar matter as a function of $n_B/n_0$. It shows a rise in temperature with $n_B/n_0$, followed by a sharp decrease at the critical temperature, $T_C$ (indicated by a black star), marking the phase transition from hadronic to quark matter phase. The core temperatures, $T_{co}$, are also highlighted with purple stars on the curves, except where they lie outside the critical temperature range.}
    \label{temp}
\end{figure}
In Fig.~\ref{temp}, we depict the temperature as a function of baryon density. The Maxwell construction of hybrid stars involves two different phases of matter; therefore, the two phases are characterized by different temperatures. This construction matches the pressure of the phases at a fixed chemical potential. Therefore, to ensure thermodynamic equilibrium and stability, the phase transition at the boundary in a hybrid star must be isothermal. 
In phase transition, the corresponding $T,\, \mu_B, P,\; {\rm and},\; \varepsilon$ for both phases are related to the values determined at $T_C=T_H=T_q$ (where $T_q$ and $T_H$ are the temperatures in the quark and hadron phases, respectively), $P_c$, and $\mu_c$ at the isothermal intersection for the fixed $S/n_B$. Determining the intermediate phase-transition quantities for the mixed phase requires knowledge of the Gibbs construction of hybrid stars \cite{Shao:2011nu, Menezes:2003pa}, which is beyond the scope of this work. The drop in temperature as the QM phase emerges is partly attributed to the high density of the medium and the melting of diquark condensates to form the QM phase (see Ref.~\cite{Carlomagno:2024vvr, Mariani:2016pcx} for similar discussions). 

Moreover, there is a possible coexistence of hadronic and QM phases at the point of transition; this increases the degrees of freedom beyond the degrees of freedom of pure hadronic matter, thereby reducing its temperature. As the density increases in the transition phase, the quarks gradually dominate the matter, and the hadron and lepton concentrations in the stellar matter begin to reduce significantly, as shown in the lower panels of Fig.~2. This further reduces the temperature; however, when the stellar matter finally converts into pure quark matter, the temperature begins to rise steadily with density again \cite{Logoteta2022}, as can be seen above for the $S/n_B=2$, $Y_{\nu_e}=0$ curves. 

Generally, we observe that an increase in the degrees of freedom of the stellar matter leads to a decrease in the net temperature of the stellar matter. Ideally, increasing the degrees of freedom increases the entropy of stellar matter. {Since entropy is kept fixed at each stage, an increase in the degrees of freedom will decrease the net temperature of the stellar matter.} Also, in the neutrino-trapped matter, we observe that the temperature profile in the first stage when the star is capturing neutrinos, 
constitutes the least profiles. 
When the star begins deleptonizing, the temperature of the stellar matter increases, as represented by the second stage in Fig.~\ref{temp} (for temperature fluctuation in PNSs with exotic baryons, see \cite{Issifu:2023qyi, Sedrakian:2022kgj, Raduta:2020fdn}). At this point, the temperature of the stellar matter is relatively higher. In the third stage, when the quark core begins to form in the stellar matter, the temperature increases until it reaches a critical point (marked with black stars on the curves) in the hadronic matter region. Subsequently, it decreases sharply as free quarks emerge in stellar matter. The sharp decrease in the temperature when the quark core starts forming is attributed to the higher degeneracies of the quarks. The core temperatures are denoted by purple stars on the curves, except where the core temperature falls outside the critical temperature range.

\begin{figure}[ht!]
  \includegraphics[scale=0.5]{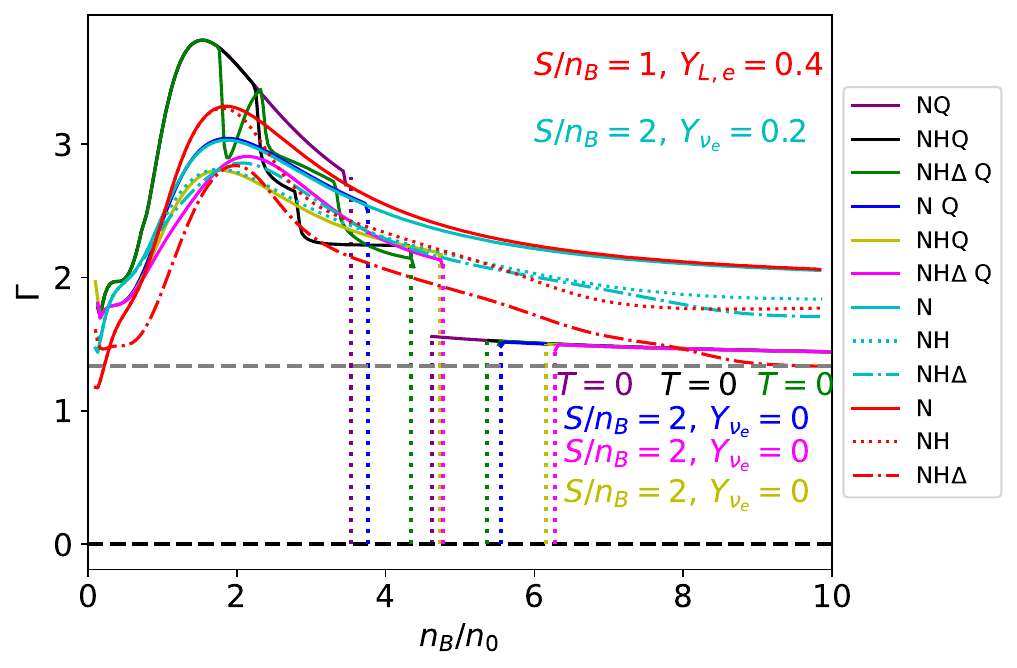}
\caption{The curves represent the adiabatic indices of the hadronic and hybrid stars as a function of $n_B/n_0$. A horizontal gray line at $\Gamma = 4/3$ denotes the instability threshold. Curves intersecting this line are deemed unstable, while those above it are considered stable. The black horizontal line is the $\Gamma =0$ line. The interval between two dotted vertical lines of the same kind that touches the $\Gamma =0$ line is the region where phase transition takes place. The red curves depicting the PNS in its early stages intersect the instability line at various densities. Both the N and NH curves cut the instability line at the low-density region while the NH$\Delta$ curve cuts it at the high-density region. }
    \label{adiab}
\end{figure}
In Fig.~\ref{adiab}, we analyze the stability of PNSs (with quark cores) using the adiabatic index as the benchmark. Based on the seminal presentations by Chandrasekhar \cite{Chandrasekhar:1964zza, PhysRevLett.12.114}, it was established that the dynamical instability of compact stars can be studied through variational methods. Mathematically, the adiabatic index ($\Gamma$) is expressed as 
\begin{equation}
    \Gamma = \dfrac{P+\varepsilon}{P}\Bigg(\dfrac{dP}{d\varepsilon}\Bigg)_{s_B},
\end{equation}
where the term in the bracket is precisely the speed of sound and $s_B=S/n_B$ is the entropy per baryon at which the $\Gamma$ is being computed. $\Gamma$ is a dimensionless quantity determined through the EoS, hence, its value depends on the stiffness or otherwise of the corresponding EoS. A star is considered stable when $\Gamma > 4/3$ in its core. Conversely, a star is deemed unstable when $\Gamma < 4/3$, indicating an impending collapse. Consequently, a critical point exists at $\Gamma = 4/3$, serving as the boundary that distinguishes between the stable and unstable regions \cite{glass1983stability, PhysRevC.50.460}. 

From Fig.~\ref{adiab}, it is evident that the PNS formed during the initial stage of stellar evolution when neutrinos are tapped in its core 
intersects the instability threshold at various 
$n_B/n_0$ values. This indicates that PNSs formed in this regime are unstable regardless of the composition of the stellar matter, as highlighted by the red curves. The rest of the stars satisfy the stability conditions set through $\Gamma$. After the neutrinos have escaped and the quark core starts forming, the resulting hybrid stars remain stable as shown in the figure above. However, at the point where the quark core starts forming, the curves of the stars decrease steadily until they approach the instability line from above but do not cross it. Thus, the stars with the quark core are stable as well.

\begin{figure*}[ht!]
  \includegraphics[scale=0.5]{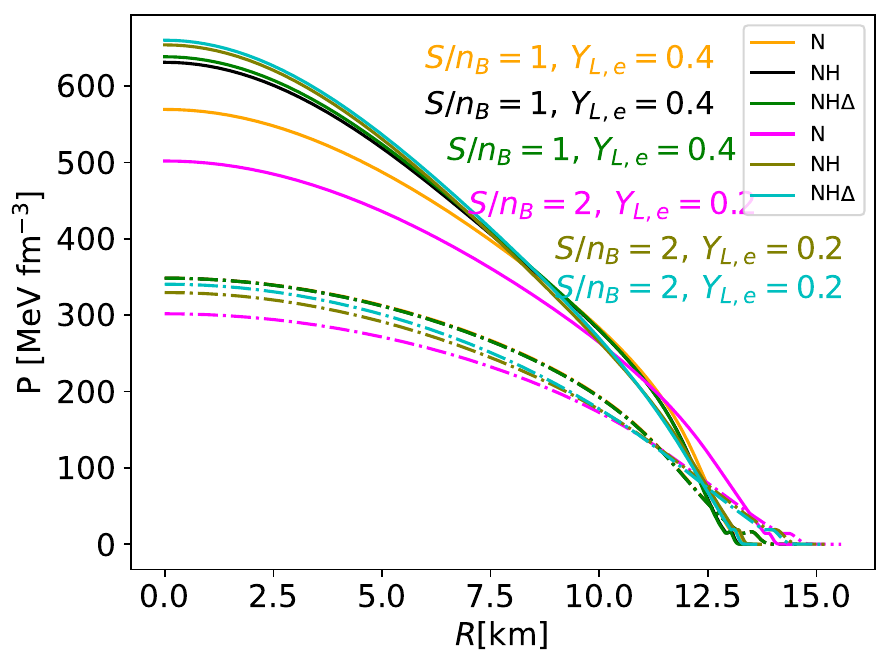}
  \quad
   \includegraphics[scale=0.5]{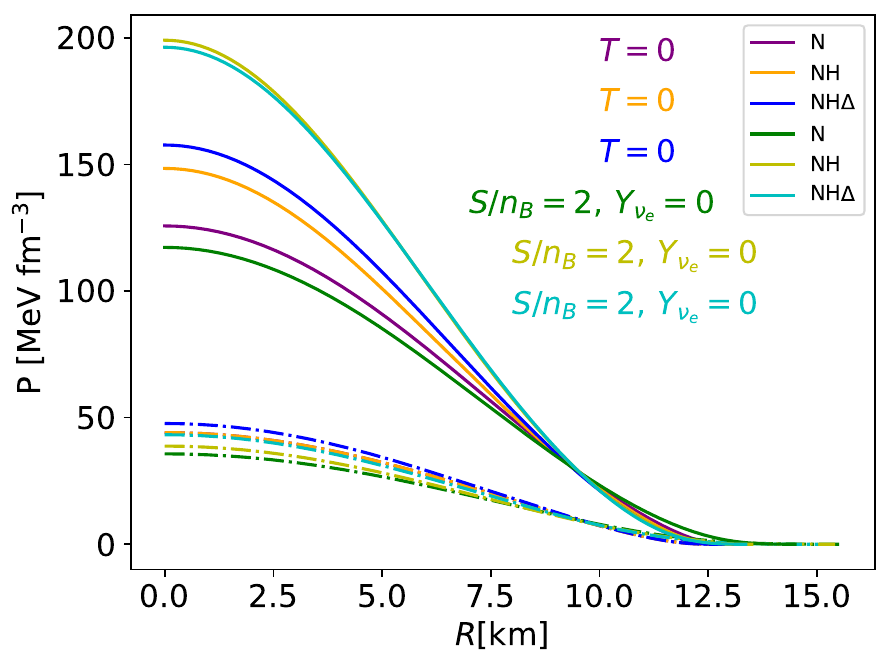}
  
\caption{The pressure profile of two stars with fixed baryon masses, $M_b=1.57\, \rm M_\odot$(dot-dash lines) and $M_b=2.45\,\rm M_\odot$(solid lines). Comparatively, the presence of trapped neutrinos in the star increases its pressure drastically, this is not immediately obvious when we consider a group of stars as in Fig.~\ref{eos} due to the corresponding increase in energy density.}
    \label{p}
\end{figure*}
Subsequently, we profile two stars with fixed baryon mass to determine how the pressure, temperature, and sound velocity behave with radius in relatively lighter and heavier stars. It is important to note, that during the stellar evolution, the baryon density does not vary significantly, as indicated by the maximum baryon masses in Tab.~\ref{T1a}. In the subsequent figures, solid lines describe a massive star with $M_b=2.45\,\rm M_\odot$, while dashed-dot lines describe the corresponding $M_b=1.57\,\rm M_\odot$ stars. In the left panel of Fig.~\ref{p}, we display the pressure profile of the neutrino-trapped matter, while in the right panel, we illustrate the pressure profile of the neutrino-transparent matter. Generally, a more massive star is associated with higher core pressure than a lighter star. However, both exhibit a constant pressure at the surface of the star. Comparing the two panels, it's evident that trapped neutrinos increase the pressure in the core of the star compared to the neutrino-free ones. Nonetheless, after the neutrinos have escaped from the star, the surface pressure remains constant for a few more kilometers before it starts rising steeper towards the core compared to the neutrino-trapped ones. 

\begin{figure*}[ht!]
  \includegraphics[scale=0.5]{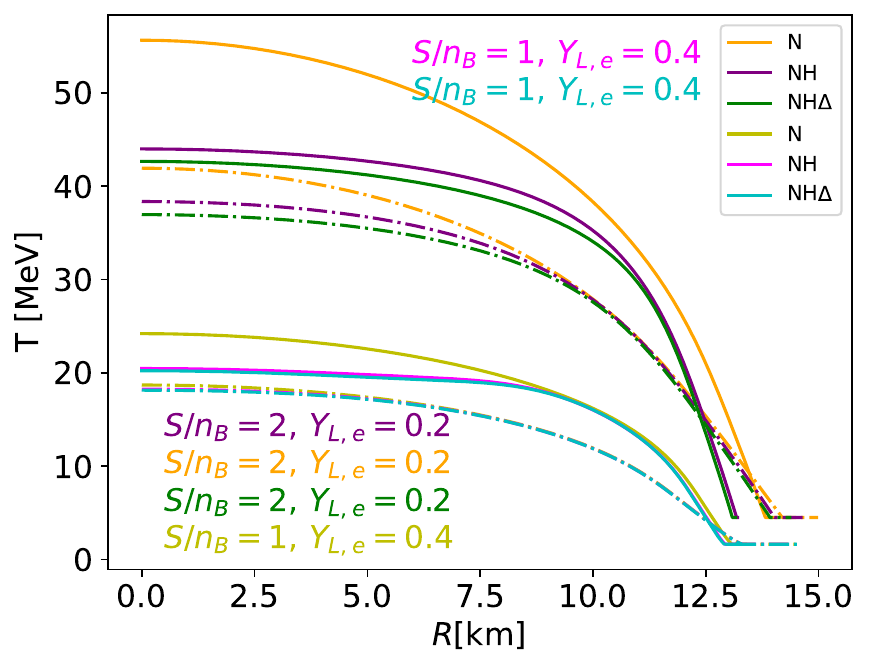}
  \quad
   \includegraphics[scale=0.5]{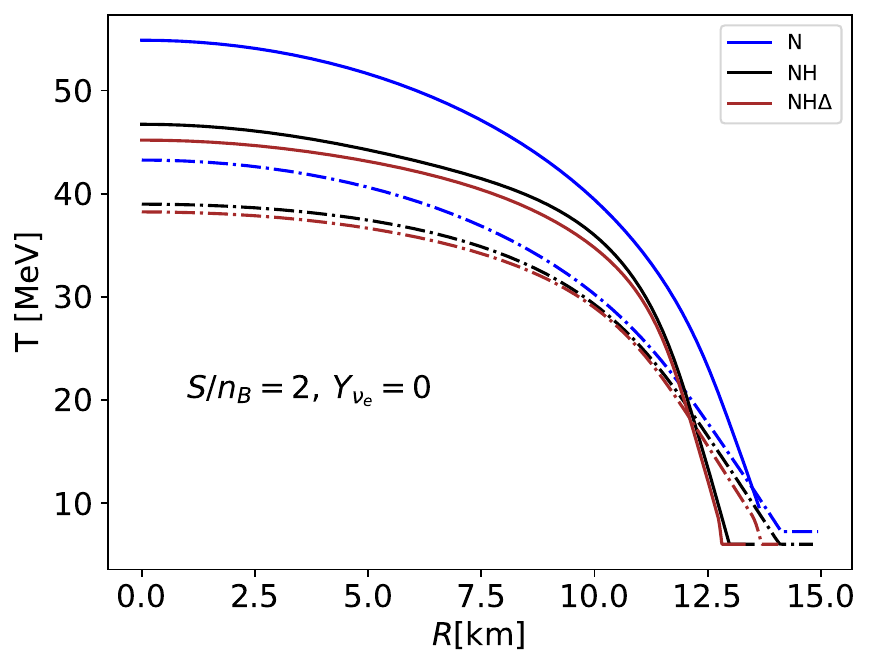}
  
\caption{The temperature profiles for the two selected stars are depicted on the graph. Dot-dashed lines distinguish the star with $M_b=1.57\, \rm M_\odot$, while the star with $M_b=2.45\, \rm M_\odot$ is represented by solid lines.
}
    \label{Tp}
\end{figure*}
In Fig.~\ref{Tp}, we present the temperature profiles in the core of the two selected stars. The left panel represents the instance where neutrinos are trapped inside the star, while the right panel illustrates when the neutrinos have all escaped from the star's core, and the stellar matter is maximally heated. Generally, the temperature is higher in more massive stars than in lighter ones, as depicted in the figure above. Additionally, the temperature remains constant for a few kilometers on the star's surface before it starts increasing towards the core of the star. The initial stages of star formation are associated with lower temperature profiles due to higher $Y_{L,e}$, as expected. As the star begins deleptonizing, the temperature of the stellar matter increases as depicted in the figure above.

\begin{figure*}[ht!]
  \includegraphics[scale=0.5]{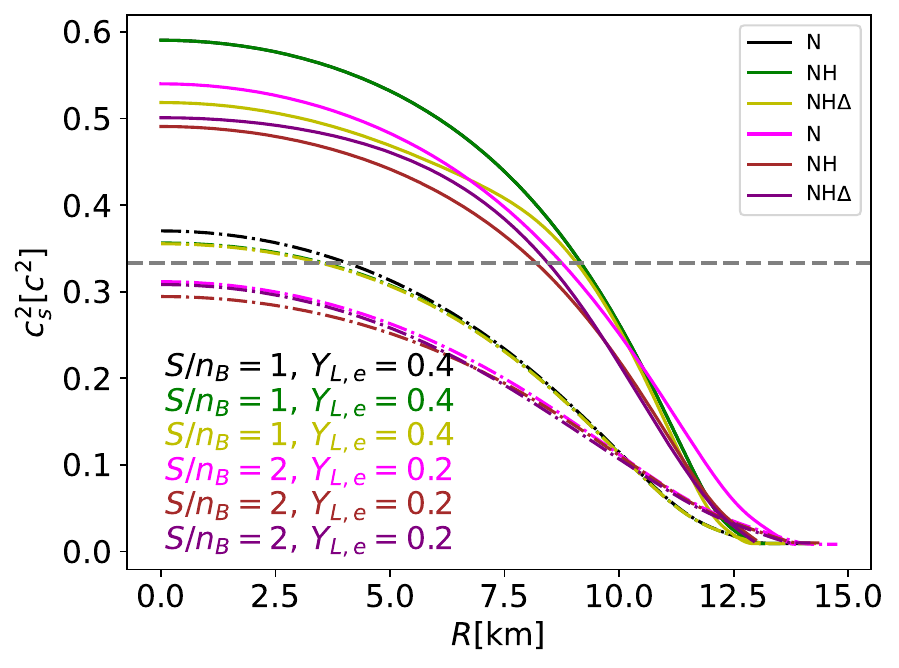}
  \quad
   \includegraphics[scale=0.5]{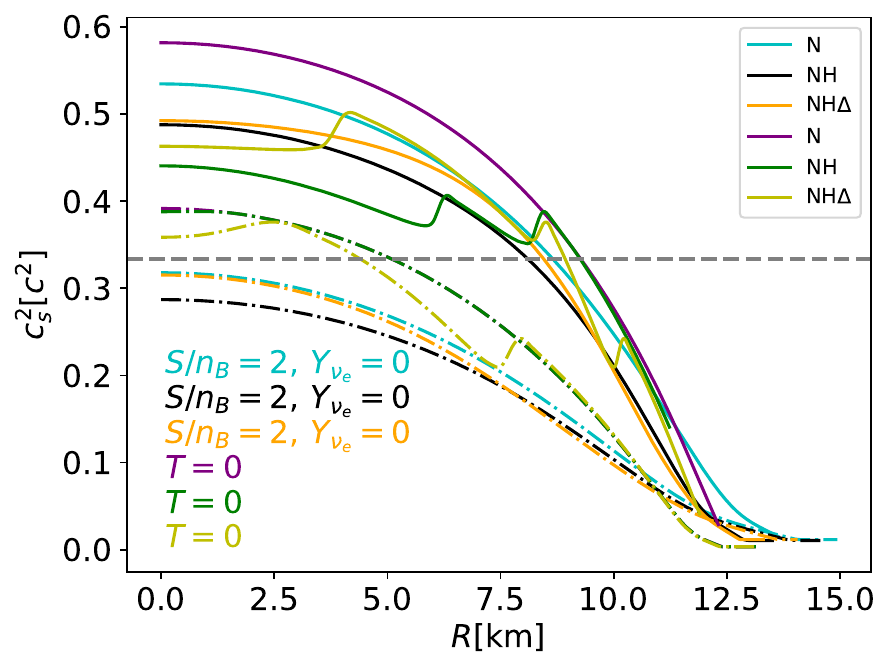}
  
\caption{The $c_s^2$ of two stars with fixed baryon masses, $\rm M_b=1.57\,\rm M_\odot$(dot-dash lines) and $\rm M_b=2.45\, \rm M_\odot$(solid lines).}
    \label{sv}
\end{figure*}

In Fig.~\ref{sv}, we illustrate the $c_s^2$ of the two selected stars as a function of the star's radius. This will allow us to determine the composition of matter that makes up these stars. The left panel represents stars with neutrinos trapped in their core, which we know 
to be composed only of hadrons. Therefore, their $c_s^2$ are expected to rise monotonously towards the core, as shown in the graph above. In the right panel, we display the results of the stars formed with neutrino-transparent matter, with a possible quark core formation. Generally, the $c_s^2$ is higher in heavier stars than in lighter ones. Also, relatively colder stars are associated with higher sound velocity than their hotter counterparts. We demarcate the conformal boundary with a horizontal gray line at $c_s^2=1/3$. In smaller and hotter stars, such as those with $S/n_B=2$ and $Y_{L,e}=0.2$, as well as those with $S/n_B=2$ and $Y_{\nu_e}=0$, the $c_s^2$ falls below the conformal limit. This might suggest the possibility of free quark phase transitions. However, upon closer examination of the $n_c$ values for $M_b=1.57\,M_\odot$ stars, it becomes evident that this behavior is not indicative of a quark phase transition at this point in time. Instead, it can be attributed to the mass of the stars under consideration. In contrast, the $c_s^2$ determined for heavier stars with $M_b=2.45\,M_\odot$ exhibits typical characteristics of hadron stars. 
As discussed below Fig.~\ref{sound}, the occurrence of the bumps in $c_s^2$ observed in the stars at $T=0$ coincides with the appearance of new degrees of freedom in the star when it is cold.

\section{Final remarks}\label{fr}
We investigated the evolution of PNSs from their birth to maturity to determine the conditions under which a quark core begins to form in their interior, amidst extreme temperatures and densities. The stellar matter comprises nucleons, nucleons plus hyperons, and nucleons plus hyperons plus $\Delta$-isobars. For the first time, we used the DDQM and RMF approximation with density-dependent coupling to show that the formation of the quark core during stellar evolution begins only when all neutrinos have escaped from the star. The study demonstrates that the presence of neutrinos in the core of the PNS hinders the formation of free quarks in its core. However, once all the neutrinos have escaped from the stellar core and the stellar matter is expected to reach maximum heating, the phase transition begins to occur. Consequently, within the model framework, the phase transition is found to be highly dependent on the lepton fractions rather than the temperature of the stellar matter. The observed drastic drop in temperature and sound velocity at the onset of the quark core can indeed serve as a reliable indicator for detecting hybrid stars, as our results demonstrate. In \cite{Sandin:2007zr, Ruester:2005ib}, the authors explore the impact of neutrino trapping on the quark-hadron phase transition in PNSs. They conclude that the appearance of a quark core in PNSs before deleptonization is completed is highly unlikely. These conclusions are consistent with our findings. Our study led to the determination of the critical chemical potential, critical temperature, and critical pressure at which the hadron-quark phase transition can occur in neutrino-transparent matter for different compositions of matter, and the results are displayed in Tab.~\ref{T1a}. 
The specific findings of the study are summarized below:
\begin{itemize}
        \item  In Fig.~\ref{equ}, we observe an overlap between the NH and NH$\Delta$ curves, even though NH$\Delta$ includes additional degrees of freedom compared to NH. This is attributed to the baryon-meson couplings, which are adjusted to yield stellar masses within the 2$\rm M_\odot$ threshold, even though hyperons and $\Delta$-isobars soften the EoS. With these couplings, the EoS is softened at lower densities and stiffened at higher densities to achieve the 2$\rm M_\odot$ threshold. However, comparing the left (neutrino-trapped matter) and the right (neutrino-transparent matter) panels, the $P-\mu_B$ graphs are nearly indistinguishable. However, the $\Delta$-isobar content is prominently visible in \( Y_i \), in Fig.~\ref{pfs}. Moreover, $c_s^2$, $T$ and $\Gamma$ plots in Figs.~\ref{sound}, \ref{temp}, and \ref{adiab}, and the stellar profiles in Figs.~\ref{p}, \ref{Tp} and \ref{sv} demonstrate differences between NH and NH$\Delta$ (particularly for the neutrino-trapped matter) at intermediate to higher $n_B$. 

      \item We calculated the $Y_i$, and presented our results in Fig.~\ref{pfs} to illustrate the various particle compositions in the stellar matter, at which density they appear in the stellar matter, and to pinpoint when the presence of free quarks become apparent in the stellar matter. We find that there is a gradual phase transition from hadronic to quark matter phase after the neutrinos have escaped from the stellar core ($S/n_B=2$ and $Y_{\nu_e}=0$) within the baryon density range $3.73\leq n_B[n_0]\leq6.28$. After the stellar matter has cooled down to $T=0$, the phase transition accelerates, and the quark core expands relatively. This transition occurs within the baryon density range $3.53\leq n_B[n_0]\leq 5.47$. Similar studies were conducted in \cite{PhysRevC.100.015803}, where the authors found phase transitions occurring solely at $T=0$. Additionally, in \cite{Shao:2011nu}, the authors conducted similar investigations using Gibbs construction for hybrid stars, considering only nucleons in their model.
 
 \item  In Fig.~\ref{eos}, we presented the EoS, which forms the foundation of the entire study. We observe that in the neutrino-trapped region, the increase in $Y_{L,e}$ results in the stiffening of the EoS, and vice versa. However, in the neutrino-transparent matter, a hadron-quark phase transition occurs, leading to a noticeable discontinuity in the energy density. Here, the EoS softens as the stellar matter cools from $S/n_B=2$ and $Y_{\nu_e}=0$ to $T=0$ through the emission of thermal radiations (see e.g. \cite{PhysRevC.100.015803, Sedrakian:2022kgj, Raduta:2020fdn, Issifu:2023ovi} for an insight into fixed entropy EoSs and hybrid EoS).

 \item In Fig.~\ref{mr}, we presented the results of the structure of the stars. We find that the maximum masses are consistent with those obtained for PSR J0740+6620 \cite{NANOGrav:2019jur, Fonseca:2021wxt}, PSR J2215+5135 \cite{Linares:2018ppq}, and PSR J0952-0607 \cite{Romani:2022jhd} pulsars. Additionally, the radius falls within the bounds of the pulsars PSR J0740+6620 and PSR J0030+0451 determined through the analysis of the NICER data. Other observable data that satisfied our results have been indicated in Fig.~\ref{mr}.
 
 \item We distinguished between the hadronic and quark matter phases \cite{Annala:2019puf, Annala:2023cwx} calculating  $c_s^2$ in Fig.~\ref{sound}. The $c_s^2$ exhibits a rise with increasing density in the hadronic regime and then begins to decrease sharply when the quark core starts to appear {touching the $c_s^2=0$ line and rising again}, eventually reaching {its maximum} value below the conformal limit. Moreover, we profiled two different stars with $\rm M_b = 1.57\,M_\odot$ and $\rm M_b = 2.45\,M_\odot$ and used the $c_s^2$ as a function of star's radii (Fig.\ref{sv}) and $n_c$ (Tab.\ref{T1a}) to determine their particle composition. We found that both stars profiled are hadron stars. Additionally, we observed that the $c_s$ in the core increases when neutrinos are trapped in the stellar matter. It drops immediately after all the neutrinos have escaped from the core and increases again when the star cools down to $T=0$.

 \item We calculated the variation of temperature with the baryon density and presented our findings in Fig.~\ref{temp}. Additionally, we studied the temperature profile in two selected stars and presented our results in Fig.~\ref{Tp}. Generally, the temperature rises with $n_B$ and $R$ towards the stellar core. However, when new degrees of freedom appear in stellar matter, they significantly reduce the temperature gradient. In Fig.~\ref{temp}, where the $S/n_B=2$ and $Y_{\nu_e}=0$ stellar matter possesses the quark core, the temperature drops drastically as the quark core begins to form in the stellar matter. This represents a hadron-quark phase transition when the hadronic matter reaches a critical temperature $T_C$. The observed properties of the temperature variations qualitatively agree with the ones presented in \cite{Issifu:2023qoo, Raduta:2020fdn, Issifu:2023ovi, Carlomagno:2024vvr, Mariani:2016pcx}.

 \item We explored the stability of both the hadronic and hybrid NSs using the adiabatic index as our benchmark for analysis. We found that the 
 neutrino-rich ($S/n_B=1$ and $Y_{L,e}=0.4$) PNS is unstable for the various particle compositions, as illustrated in Fig.~\ref{adiab}. The star stabilizes during deleptonization, transitioning through neutrino transparency to the formation of a cold-catalyzed hybrid NS. The point of the hadron-quark phase transition is depicted by a steep drop in the curves {touching the $\Gamma =0$ line before rising again beyond the instability threshold to near constant, approaching the $\Gamma=4/3$ from above}. A detailed discussion on the derivation of $\Gamma$ and the conditions under which stability can be determined within its framework is presented in \cite{Moustakidis:2016ndw}.

 \item We profiled the pressure variation as a function of the star's radii of two NSs with fixed baryon masses as shown in Fig.~\ref{Tp}. We observed that the pressure is relatively constant on the star's surface for both the lighter and the heavier stars analyzed, and it begins to rise with radius towards the star's core a few kilometers from the surface. The results indicate that the presence of neutrinos in the star's core significantly increases its core pressure. As the star undergoes deleptonization and neutrino diffusion, the pressure begins to drop and then rises again when the star catalyzes, shrinks, and compactifies to form a  ``mature" NS at $T=0$.
 
 \end{itemize}
From the study, we determined that the presence of neutrino concentration in the stellar core during its formative stages impedes the dissociation of the hadrons into SQM in the stellar core. Additionally, neutrino concentration suppresses temperature rise, increasing pressure and $c_s^2$ of the stellar matter toward the star's core. This results in an increase in the maximum mass of the neutrino-trapped PNSs relative to neutrino-poor ones. During the deleptonization stages, the matter heats up, leading to an expansion in the size of the star and a relative drop in the maximum gravitational mass, as shown in Tab.~\ref{T1a}, in the absence of accretion and the formation of a black hole. We established the formation of a quark core when all the neutrinos had escaped from the stellar core. {It would be nice to reproduce the present study with other models to check how model-dependent our conclusions are.} 

\section*{Acknowledgements}
This work is a part of the project INCT-FNA Proc. No. 464898/2014-5. D.P.M. was partially supported by Conselho Nacional de Desenvolvimento Científico e Tecnológico (CNPq/Brazil) under grant 303490-2021-7. A.I. thanks the  financial support from the
 S\~ao Paulo State Research Foundation (FAPESP) Grant No.  2023/09545-1. T. F. thanks
the financial support from the Brazilian Institutions: CNPq (Grant No. 306834/2022-7), Improvement of Higher Education Personnel CAPES (Finance Code 001) and FAPESP (Grants 
No. 2017/05660-0 and 2019/07767-1). 
Z.R. wishes to thank the Shiraz University Research Council.

\bibliography{referens.bib}

\end{document}